\documentclass[twocolumn]{article}
\usepackage[numbers,sort&compress]{natbib}
\usepackage[utf8]{inputenc}
\usepackage{amsmath}
\usepackage{amsfonts}
\usepackage{graphicx}
\usepackage{tabularx, ragged2e, booktabs}
\usepackage{algorithm}
\usepackage[noend]{algpseudocode}
\usepackage{xcolor}
\usepackage{subfigure}
\usepackage{tikz,tikz-inet}
\usetikzlibrary{arrows}
\usepackage{multirow}
\usepackage{setspace}
\usepackage{adjustbox}
\usepackage{balance}
\usepackage{units}
\usepackage{blindtext}
\usepackage{mathtools}
\usepackage{url} 
\usepackage{authblk}

\newtheorem{definition}{Definition}[section]

\title{A Centrality Approach to Select Offloading Data Aggregation Points in Vehicular Sensor Networks}

\author[1]{Douglas Moura}
\author[2]{Geymerson S. Ramos}
\author[3]{Andre L. L. Aquino}
\author[1]{Antonio Loureiro}

\affil[1]{Department
of Computer Science, Federal University of Minas Gerais, \newline Belo Horizonte, MG, Brazil}
\affil[2]{INSA Lyon, Villeurbanne, France}
\affil[3]{Computer Institute, Federal University of Alagoas, Maceio, AL, Brazil}

\date{}

\begin{document}

\maketitle

\begin{abstract}
This work proposes a centrality-based approach to identify data offloading points in a VSN. 
The solution presents a scheme to select vehicles used as aggregation points to collect and aggregate other vehicles' data before uploading it to processing stations.
We evaluate the proposed solution in a realistic simulation scenario derived from data traffic containing more than 700,000 individual car trips for 24 hours.
We compare our approach with both a reservation-based algorithm and the optimal solution.
Our results indicate an upload cost reduction of 30.92\% using the centrality-based algorithm and improving the aggregation rate by up to 10.45\% when considering the centralized scenario.

\end{abstract}

\noindent \textbf{Keywords:} Offloading, VANETs, Cellular networks, simulation

\section{Introduction}

A Vehicular Ad Hoc Network (VANET) presents different vehicles acting as mobile sensors~\citep{ISCC2018:Rettore+5}, which allow monitoring of urban environments, provide efficient traffic management, and realize long-term urban planning~\citep{wang2018}.
A Vehicular Sensor Network (VSN) is a fusion of a VANET and a Wireless Sensor Network (WSN) without the power restrictions typically in a sensor network.
On the other hand, a VSN might have powerful processing units, wireless communication, GPS receivers, and a plethora of sensing devices~\citep{fadi2020}. With a vehicular sensor network, we can design more sophisticated applications for road safety, traffic management, intelligent navigation, pollution monitoring, urban surveillance, and forensic investigations \citep{bazzi2015}. Some data can be used directly by applications. However, artificial intelligence and machine learning bring unprecedented opportunities to handle and analyze these data. We can extract more information and build models of complicated systems, such as citizen behaviors, to understand the city comprehensively and even predict the dynamics of systems~\cite{du2018sensable}.

Many continuous monitoring applications require periodic data uploading, creating significant traffic in the uplink channel. 
According to Cisco~\cite{cisco}, monthly mobile data traffic predictions will reach 77 exabytes by 2022, an increase of more than 165\% compared with 2019, and 20\% of IP traffic will be from mobile devices. 
The fifth-generation network (5G) is the technology to support this massive data growth, consolidate the Internet of Things paradigm, and connect different wireless devices. 
Although this scenario favors new VANETs based on cellular communication, the rapid traffic growth requires reducing the overhead in the cellular network.
Different studies~\cite{Stanica2013, bazzi2015} consider offloading mobile data to overcome capacity and cost challenges. 
In this scenario, mobile data offloading consists of transferring cellular traffic to alternative networks to reduce the traffic load in the cellular network.
Some proposals transmit offloaded cellular data in an uplink channel to complementary and low-cost networks, such as Wi-Fi and IEEE 802.11p, or using a Device-to-Device (D2D) protocol~\cite{mao2016}. 
In addition to alternative infrastructure networks, some vehicles can act as aggregation points. Local aggregation techniques can combine sensory data from different sources, eliminate redundancy, and reduce the number of transmissions.

This work focuses on periodic sensor data uploading for analysis and remote processing. 
In this context, the sensing application originates massive data and transmits it through a cellular network, requiring strategies to save bandwidth and prevent overload.
We explore the structural network information to perform the data offloading in a vehicle sensor network and reduce the volume of sensor data uploaded in the cellular network.
Our main research question is:
    \textit{``How to select a subset of vehicles to receive, aggregate, and forward sensor data offloaded through D2D communication?''}

D2D communication has emerged as a promising technology to allow spatial frequency reuse of cellular networks.
This way, VSN can use the D2D communication to offload sensory data with higher data rates, low latency, and low-energy consumption through a direct link between nearby vehicles.
Our approach aims to use the minimum number of vehicles transmitting in a cellular network and reduce the upload cost.
To determine this minimal number of vehicles, we model the problem as a variation of the dominating set problem and present a solution based on the closeness centrality measure~\cite{FREEMAN1978215} to find a set of aggregation vehicles that minimize the uplink traffic volume.

The main contribution of this work is the proposal of a new heuristic to select aggregation points responsible for performing data offloading in a centralized or decentralized manner in a vehicular sensor network.
Unlike previous results, our solution does not need to deploy a complementary communication infrastructure or synchronized transmissions between vehicles~\cite{Stanica2013, bazzi2015,Lin2018}.
Our experiments consider different evaluation metrics and offloading conditions. 
We use a realistic scenario obtained from synthetic mobility traces. 
The results significantly reduce bandwidth usage compared to the traditional upload solution and another offloading approach. 
Additionally, we observe that the centrality metric can improve the selection of aggregation points in terms of cluster stability, even using multi-hop communication.


\section{Related work}\label{sec:related}

Mobile data traffic will expand very quickly in the next years~\cite{cisco}, calling for new solutions to the traffic overhead problem in a scenario of a vehicular sensor network.
In addition to energy savings, data aggregation techniques can significantly enhance network throughput in a VSN~\citep{Krishnamachari}.
Moreover, packet transmissions in the cellular network can be monetarily expensive and inefficient in satisfying mobile users' demands~\citep{Lin2018}.
In this case, devices may use additional network infrastructure with a higher capacity to share collective interest content in an ad-hoc way.
However, some features of a vehicular sensor network, such as a highly dynamic mobility pattern, require specific approaches when dealing with these issues.

In recent years, many proposals focused on offloading techniques to reduce the load on cellular networks. However, a fundamental problem in Multi-Access Edge Computing (MEC) offloading is selecting where to perform the offloading. In~\cite{min2019}, the authors addressed this problem and proposed two learning-based task offloading schemes for IoT devices with energy harvesting. The results showed reduced energy consumption, computation latency, and task drop rate. Although their solution explores the offloading concept, they only considered task offloading.

Kumar et al.~\cite{kumar2015} proposed a collaborative learning automata-based routing for rescue operations using VSNs. In their work, vehicles collaborate to share sensor data and intelligently select the best route to reach the final destination. Kumar et al.~\cite{kumar2013} used a learning-based clustering to determine the cluster head based on the direction of mobility and density of vehicles. Kumar et al.~\cite{kumar2015bayesian} formulated the problem of reliable data forwarding as a Bayesian Coalition Game (BCG). They proposed a new algorithm called Learning Automata-based Contention Aware Data Forwarding (LACADF) for critical Vehicular Cloud (VCloud) applications. Similarly, Singh et al.~\cite{Singh} used Ant Colony Optimization (ACO) to propose a routing algorithm based on the random selection of source and destination nodes. Although the studies mentioned above present the feasibility of several techniques, their scope is limited to routing and information sharing.

Studies about offloading to improve traffic downlinks mainly focus on disseminating and downloading user content.
Mezghani et al.~\cite{mezghani2016} presented an innovative seed selection scheme, SIEVE, to offload popular content from the cellular network through the VANET network according to user preferences. The simulation results showed that SIEVE could achieve a higher coverage rate, around 89\%, in a low-density scenario.
Dua et al.~\cite{Dua2017} considered game theory to calculate the utility of the nearby vehicles and Wi-Fi access points to act like players in a game and perform mobile data offloading.
Similarly, Mao et al.~\cite{mao2016} proposed to increase the communication capacity of a cellular network by using D2D communication and high-capacity Wi-Fi networks, integrating different devices in a heterogeneous network.
Both studies used Wi-Fi networks as an alternate way to disseminate data to the recipient.
Data offloading in a vehicular sensor network have requirements that differ from downlink-based offloading. The offload must be scalable and support a large-scale network with millions of sensor vehicles generating massive data to be uploaded. Also, the offload must be reliable and sensitive data must be reported to the monitoring center without loss.
Previous solutions typically address delay-tolerant approaches, while many VSN-based services require near real-time upload as real-time traffic control.

From a different perspective, some studies consider mobile data offloading focused on uplink data transfers, which is also the central subject of this work.
Stanica et al.~\cite{Stanica2013} combined data aggregation techniques with mobile data offloading to reduce the number of transmissions in the cellular uplink.
That work proposed three distributed algorithms to allow massive data offloading: Degree-Based (DB), Degree-Based with Confirmation (DB-C), and, finally, Reservation-Based (RB), which presented the best results when a complete coverage of the vehicles is required.
The RB algorithm showed a system gain of approximately 80\% at peak hours, but the synchronization requirements limit its usage.
We compare the RB algorithm with our solution once it presents a similar strategy and the most promising results in the literature. 

Some proposals use opportunistic communication to offload delay-tolerant data~\citep{tamai2018}. 
A disadvantage of these approaches is that content offloading will cause a longer delay than direct transmission in cellular networks.
When we consider complementary network infrastructures, such as IEEE 802.11p Roadside Units (RSUs)~\cite{bazzi2015,raja} and Wi-Fi access points to offload data~\cite{Lin2018}, it is possible to achieve a significant reduction of overload in the cellular network.
However, the opportunistic behavior of these approaches is only compatible with some applications that require a fine granularity of the data acquisition w.r.t. time and space.

Finally, some solutions use centrality metrics in VANETs and data dissemination. 
Moura et al.~\cite{moura} used the betweenness centrality to identify the dissemination points.
As in this work, the authors used a centrality measure to identify nodes, which due to their topological position in the network, are more influential in the flow of information.
Other studies explore graph theory to extract structural properties from the network and perform the offloading.
Yan et al.~\cite{yan2016} used graphs to model vehicle connections' probability. 
That work proposed a space and time-constrained data offloading scheme using a probability-based contact graph to describe near-term transmission opportunities among vehicles. Their results showed an increase of more than 70\% in data offloading in a time-space-constrained scenario.

Our work differs from the previous one since we do not deploy complementary infrastructure to the cellular network nor use a synchronization process during the transmissions. 
Our solution can perform the offloading even in near real-time sensing and operates decentralized. 
We identify other relevant improvements (Section \ref{sec:experiments}) when we compare our approach with the RB algorithm~\cite{Stanica2013} mentioned above.

\section{Offloading data aggregation problem}\label{sec:problem}

\begin{figure}[htb]
\centering
\subfigure[Data collection]{\includegraphics[width=.31\columnwidth]{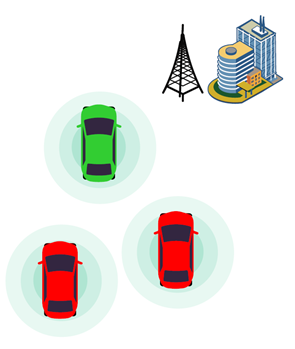}}
\hspace*{\fill}
\subfigure[Data aggregation]{\includegraphics[width=.31\columnwidth]{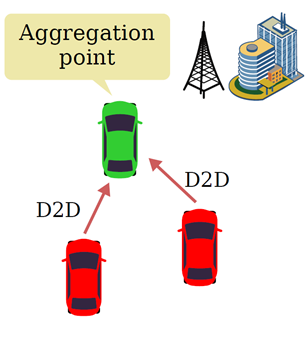}}
\hspace*{\fill}
\subfigure[Data delivery]{\includegraphics[width=.31\columnwidth]{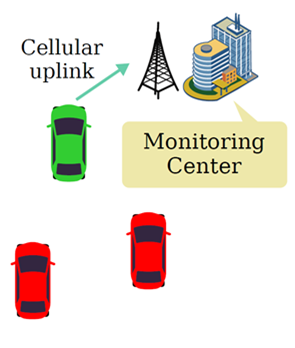}}
\caption{Mobile data offloading in vehicular sensor networks.}
\label{fig:schema}
\end{figure}

The general data offloading process in a VSN is depicted in Figure \ref{fig:schema} and presents three main steps: 
(i) \textit{Data collection\/}: each vehicle acts as a mobile sensor and performs urban sensing. 
The offloading of collected data occurs to some neighboring vehicle (called aggregation point) through D2D connections;
(ii) \textit{Data aggregation\/}: the data generated by adjacent sensors are often redundant and highly correlated.
Therefore, reduction and data compression techniques may eliminate redundancy and reduce the data load of a cellular network~\cite{vasconcelos2018data}.
We can use the aggregation operators (e.g., maximum, minimum, average) to combine data from several sources into a single value.
Determining the optimal aggregation points is also an NP-Hard problem~\cite{Amis}. Thus, we need to propose approximate algorithms to find the set of aggregation points that minimize the cellular network cost; 
(iii) \textit{Data delivery\/}: the aggregation points upload the reduced data to the monitoring center over the cellular network.

We focus on the \textit{Data aggregation} step, precisely determining the aggregation points using a D2D communication approach.
We model the problem as an instance of the \textit{minimum d-hop dominating set problem}~\citep{Amis}, an NP-hard problem.
Thus, we propose a heuristic algorithm to find an approximate solution. Moreover, the heuristic can operate in a centralized or decentralized manner and does not require synchronized transmissions to determine the aggregation points. Besides, the centrality measure allows the discovery of aggregation points that minimize the distance for all vehicles in the neighborhood, adapted according to traffic conditions.

In a VSN, the network topology changes periodically according to traffic conditions; thus, the aggregation points change continuously over time. 
We can model this dynamic behavior as an instantaneous graph, where we can analyze each sampled graph separately and determine the aggregation points for each collection period.
Let $G(\mathbb{V}, \mathbb{E})$ be an undirected graph at a generic time instant, in which $\mathbb{V} = \{v_1, v_2, ..., v_n\}$ denotes the set of moving vehicles, and $\mathbb{E}$ represents the set of edges. 
If $e_{i, j} = \{v_i, v_j \} \in \mathbb{E}$, then $v_i, v_j \in \mathbb{V}$ are adjacent, neighbors, and within the transmission range each other.

The $1$-hop neighborhood $N_1(v_i) = \{v_j \mid e_{i,j} \in \mathbb{E}\}$ consists of the set of all direct neighbors of $v_i$. That is, any nearby vehicle is directly reachable via D2D communication.
Therefore, the $d$-hop neighborhood 
\begin{align}
    N_d(v_i) &= \{v_j \mid \text{dist}(v_i, v_j) \leq d\}\label{equ:neighborhood}
\end{align}
consists of all neighbors of $v_i$ separated by at most $d$-hops, where $\text{dist}(v_i, v_j)$ is the length of the shortest path between $v_i$ and $v_j$. 
The path length between two vertices is the number of hops, i.e., the number of edges separating them.
The discovery of the neighborhood of a node is a crucial process for multi-hop communication. 
Based on this concept, we propose a heuristic to identify a set of vehicles, called aggregation points, to collect, aggregate, and transmit the neighbors' data through the cellular uplink.

As stated before, we model the VSN aggregation points problem as an instance of the \textit{minimum d-hop dominating set problem}, which can
be formulated as follows:
\begin{definition}
\label{pblm:aggregation_prob}
(Problem). Is it possible to find a subset $S \subseteq \mathbb{V} \mid \forall v_j \in (\mathbb{V} - S), \exists v_i \in S \mid 1 \leq \text{dist}(v_i, v_j) \leq d$?
\end{definition}

We are interested in finding the \textit{minimum d-hop dominating set} $S \subseteq V$ such that every vehicle $v_{j}$ not in $S$ has at least a neighbor $v_i \in S$ separated by at most $d$-hops. Thus, we define the decision variable 
\begin{align}\label{equ:dec_var}
y_i &=
\begin{cases}
    1, & \mbox{if } v_i \in S\\
    0, & \mbox{otherwise}
\end{cases}
\end{align}
indicating whether a vehicle $v_i$ belongs to a \textit{d-hop dominating set} $S$. Let us introduce an auxiliary set $W_i$ $(i = 1, 2, ..., n)$. Let $W_i = \{y_j \mid v_j \in N_d(v_i)\}$ be a set of decision variables $y_j$ for each neighboring vehicle $v_j$ at most $d$-hops away from a vehicle $v_i$.
To minimize the number of aggregation points, we consider the objective function 
\begin{align}
&\text{Min}\, \sum_{i=1}^{n} y_i\label{equa:min}\\
\shortintertext{subject to}
&y_i + \sum_{y_j \in W_i} y_j \geq 1 &\forall~i \in \{1, 2, \ldots, n\} \label{equ:minConstrain1}\\
&y_{i} \in \{0,1\}  &\forall~i~\in \{1, 2, \ldots, n\} \label{equ:minConstrain3}
\end{align}
Note that in constraint \eqref{equa:min}, we want to minimize the number of aggregation points in set $S$, where constraint~\eqref{equ:minConstrain1} guarantees that if $v_i \in \mathbb{V}$ is not in $S$, then there is at least a neighbor $v_j \in N_d(v_i)$ contained in $S$; and constraint~\eqref{equ:minConstrain3} states that $y_i$ is a binary variable.

\section{Approaches to determine offloading data aggregation points}\label{sec:solution}

The high mobility of the vehicles imposes severe challenges for data offloading, such as the constant fragmentation of the network and highly dynamic topology. We can deploy RSUs to solve connectivity problems, but deployment costs are high.
Other proposals use opportunistic communication to offload delay-tolerant data. However, a disadvantage of these approaches is that content offloading will cause a longer delay than direct transmission in cellular networks.
        
Cellular systems provide high downlink and uplink rates even in a high-mobility environment. Therefore, cellular networks are feasible communication technologies to deliver data from moving vehicles. Each vehicle will transmit sensory data through the cellular infrastructure in the traditional upload. This approach enables data acquisition with low latency but with a significant increase in cellular traffic due to the typically high frequency of the monitoring services. Thus, a more sophisticated solution, such as mobile data offloading, is required to perform sensory data acquisition.

In the following, we will discuss some approaches to offloading data aggregation points and present our solution, a centrality-based algorithm. These approaches must allow massive data offloading and support near real-time sensing. More specifically, our solution must consider the following restrictions: (i) the application originates a massive amount of sensing data, (ii) a scenario without a complementary infrastructure, (iii) an adaptive approach to lead with different latency restrictions, and (iv) sensory data within close space-time proximity has a significant correlation among them.

\subsection{Optimal Solution}\label{sec:optimal}

We model the selection of the offloading data aggregation points as a domination problem~\cite{6814727} in which a subset of vehicles will collect and aggregate data generated by direct neighbor vehicles before uploading them through the cellular network.
In the previous section, we introduced an Integer Linear Programming (ILP) formulation for a variant of the dominating set problem, called the \textit{minimum d-hop dominating set problem}. The objective is to minimize the number of aggregation points, where each node is at most $d$ hops from a node in the dominating set. We used the IBM ILOG CPLEX Optimization~\cite{cplex} to solve it. CPLEX uses a branch-and-cut procedure to build a search tree consisting of subproblems to be processed. Then, each subproblem in the tree is solved until no more active subproblems are available or exceed some limit. This approach can provide an optimal solution, but it is inefficient for large graphs with thousands or millions of vertices~\cite{pardalos2002}.
Therefore, we propose approximative methods to find an acceptable solution in polynomial running time~\cite{talbi2009}.

\subsection{Reservation-Based Algorithm}\label{sec:rb}

A well-known heuristic to determine offloading data aggregation points is the RB algorithm~\cite{Stanica2013}.
A reservation-based algorithm executes in a distributed manner in each vehicle and selects aggregation points for data offloading.
The RB algorithm uses synchronized transmissions between vehicles to determine the aggregation points. It follows three basic steps:
(i) at the beginning of each data collection process (reservation phase), each vehicle randomly selects a time slot among the $T$ available ones and enters the \texttt{contender} state;
(ii) a vehicle in the contender state waits for the chosen time slot (back-off). 
This vehicle will transmit a reservation message and enter the \texttt{dominator} state; and
(iii) a vehicle in a contender state that has received a reservation message from some neighbor cancels its back-off and changes its state to \texttt{dominated} state. Only vehicles in the \texttt{contender} state transmits reservation messages and become \texttt{dominators}; they form the subset of vehicles that act as aggregation points.
This solution's challenge is ensuring that every vehicle transmits in a different time slot in a real scenario. 
Furthermore, if a vehicle receives multiple reservation messages simultaneously, it may not decode them.

Algorithm \ref{alg:rb} presents the steps of the RB algorithm.
It receives as input the $T$ time slots available.
Before the data aggregation phase, each vehicle chooses a random time slot to transmit and waits for a reservation message. 
If a vehicle in the \texttt{contender} state receives a reservation message, it enters the \texttt{dominated} state.
In contrast, a vehicle changes its state to \texttt{dominator} when it transmits a reservation message. 

\begin{algorithm}[!htb]
\begin{small}
\begin{algorithmic}[1]
    \Require{$T$}
    \State $t \gets $ Choose the time among $T$ available slots
    \State \textit{state} $\gets $ \texttt{contender}
    \While{\textit{state} $=$ \texttt{contender}}
        \ state Wait for the chosen transmission slot
        \If{\textit{slot} $= t$}
            \ state Transmit a reservation message
            \State \textit{state} $\gets $ \texttt{dominator}
            \State \textbf{break}
        \EndIf
        \If{Received a reservation message}
            \State \textit{state} $\gets$ \texttt{dominated}
        \EndIf
    \EndWhile
\end{algorithmic}
\end{small}
\caption{RB algorithm}\label{alg:rb}
\end{algorithm}

The RB is an approximation algorithm for the minimum dominating set (MDS) problem, in which the group of dominator vehicles represents the aggregation points. The dominated vehicles offload their data to their dominator neighbors using inter-vehicular communications.
The solution results in 100\% coverage but does not guarantee the optimal solution for the MDS.
The number of vehicles and available time slots limit the RB algorithm's time complexity. The while loop (Lines 3-10) runs until the vehicle leaves the \texttt{contender} state. In the worst case, the vehicle chooses the last slot to transmit and has not received any reservation messages. Each vehicle executes the RB with a time complexity of $\mathcal{O}(T)$, resulting in $\mathcal{O}(nT)$.

\subsection{Centrality-Based Algorithm}\label{sec:centrality}

To improve the stability of the network and reduce the data traffic, we propose a new heuristic based on closeness centrality.
General vehicular networks have structures and characteristics studied through models and metrics based on graph theory. 
The centrality aims to classify a vertex (vehicle in our case) according to its relative position in the network. 
We use a centrality metric to determine which vehicles play the highest topological importance within that structure.
Specifically, the \textit{closeness centrality} of a node $v_i$ in the network is the measure of centrality defined by 
\begin{align}
    C(v_i) &= \frac{1}{\sum\limits_{v_i \neq v_j}{\text{dist}(v_i, v_j)}}.
    \label{equa:close}
\end{align}
The distance $dist(v_i, v_j)$ is the shortest path length between $v_i$ and $v_j$. The \textit{farness} of $v_i$ is the sum of all distances from $v_i$ to any $v_j$, such that $i \neq j$. Hence, the closeness centrality is the reciprocal of the \textit{farness}.

The shortest path between two vertices (vehicles) in an unweighted graph $G$ is the minor sequence of edges connecting the vertices.
We can use \textbf{Breadth-First Search} (BFS) to find the shortest paths in linear time complexity in the size of the adjacency list representation of $G$. We scanned each vertex's adjacency list when we dequeued the vertex. Since the adjacency lists represent the edges of a graph, the sum of the lengths of all adjacency lists is $\mathcal{O}(m)$. We calculate in each vertex of the graph; thus, the overall process and worst-case scenario require a time complexity of $\mathcal{O}(n m)$ to compute the closeness centrality of a network with $n$ vertices and $m$ edges.
Our primary interest is not the centrality's numerical value but the node's relative importance. 
This interest implies an additional $\mathcal{O}(n\log n)$ time cost to sort the vertices according to their centrality measures.
Here, we use BFS to find the shortest paths between vehicles regarding the number of hops. Moreover, the centrality-based algorithm allows other shortest-path algorithms, such as Dijkstra's algorithm, for weighted graphs. This strategy is proper when using key link metrics (e.g., latency, jitter, lifetime, and more) as vehicle communication costs.

Due to the changes in the network topology, we compute the aggregation points periodically, which increases the execution costs. 
Although the algorithm is easily parallelizable by running each vertex's process in different threads, it is still costly to run in large-scale networks, i.e., networks formed by thousands or millions of nodes. 
We reduce the algorithm's run time by estimating the \textit{k-closeness centrality} defined by
\begin{align}
    C_k(v_i) &= \frac{1}{\sum\limits_{v_i \neq v_j}{\text{dist}(v_i,v_j)}}, \forall\, 1 \leq \text{dist}(v_i, v_j) \leq k.
    \label{equa:close2}
\end{align}

The cutoff parameter $k$ sets a depth threshold for the search, which stops after reaching the $k$-\textit{th} level. 
The centrality calculation of vertex $v_i$ only considers the vertices separated by at most $k$-hops. 
We can describe the time cost as the number of enqueueing attempts during the BFS. Vehicular networks are highly assortative, with a correlation between the degree of a vehicle and the average degree of its one-hop neighbors~\cite{naboulsi2013}.
For a regular $\lambda$-degree graph (i.e., a graph where the degree of each vertex is $\lambda$), the cost to estimate $C_k$ at each node is $1 + \lambda + \lambda^2 + \ldots + \lambda^k = \mathcal{O}(\lambda^k)$, resulting in the final cost $\mathcal{O}(n \, \lambda^k)$.


Based on these concepts, the main idea of our \textbf{Centrality-based algorithm} (Algorithm \ref{alg:offloading}) is to explore the topology information from the network to choose the aggregation points to aggregate data offloaded by neighboring vehicles. First, the greedy algorithm calculates the $k$-closeness centrality for each vehicle. Then, it selects those vehicles that, due to their proximity to the others, have the potential to aggregate a more significant amount of data. Finally, it will combine them using some aggregation operator, such as the average function, and transmit the resulting data on the cellular uplink.
The algorithm receives the graph $G$, representing the D2D communication, the number of hops for multi-hop communication ($d$), and the maximum path length ($k$) for centrality estimation.

\begin{algorithm}[!htb]
\begin{small}
\begin{algorithmic}[1]
    \Require{$G$, $d$, $k$}
    \State $S \gets \emptyset$
    \State $C_k \gets $ Compute Equation~\ref{equa:close2} $\forall v_i \in \mathbb{V}$
    \While{$\mathbb{V} \neq \emptyset$ }
        \State Select $v_i \in \mathbb{V}$ that maximizes $C_k$
        \State $S \gets S \cup \{v_i\}$
        \For{\textbf{each} $v_j \in N_d(v_i)$}
            \If{$v_j \not \in S$}
                \State $\mathbb{V} \gets \mathbb{V} \setminus \{v_j\}$
            \EndIf
        \EndFor
        \State $\mathbb{V} \gets \mathbb{V} \setminus \{v_i\}$
    \EndWhile
    \State \Return{$S$}
\end{algorithmic}
\end{small}
\caption{Centrality-based algorithm}\label{alg:offloading}
\end{algorithm}

Algorithm \ref{alg:offloading} performs the following steps:

\begin{itemize}
    \item Line 1: creates the set $S$ of aggregation points as an empty set;
    \item Line 2: computes the closeness centrality for each vehicle in $G$ according to the cutoff $k$;
    \item Lines 3-5: removes from $G$ the most central vertex and adds it to set $S$, while there are vehicles in $G$;
    \item Lines 6-8: removes from $G$ each vertex that is a $d$-hop neighbor from the vehicle selected in \textbf{Line 4}, which is not in set $S$. This step ensures that aggregation points will be separated from each other by at least $d$-hops;
    \item Line 10: returns $S$ with the set of aggregation points when the algorithm achieves the termination condition. When the graph network is not connected, each isolated vehicle is added in $S$ and transmits its sensory data.
\end{itemize}

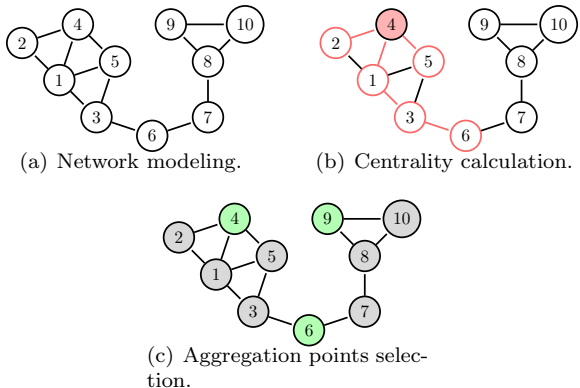
\begin{figure}[htb]
    \begin{center}
\begin{tabular}{cc}
        \subfigure[Network modeling.]{
\resizebox{0.45\linewidth}{!}{
            \begin{tikzpicture}[-,>=stealth',shorten >=1pt,auto,node distance=3cm,scale=0.6,
                    thick,main node/.style={circle,draw}]

  \node[main node, scale=0.8] (1) at (0,0) {1};
  \node[main node, scale=0.8] (2) at (-1,1) {2};
  \node[main node, scale=0.8] (3) at (1,-1) {3};
  \node[main node, scale=0.8] (4) at (0.5,1.5) {4};
  \node[main node, scale=0.8] (5) at (1.5,0.5) {5};
  \node[main node, scale=0.8] (6) at (2.5,-1.5) {6};
  \node[main node, scale=0.8] (7) at (4,-1) {7};
  \node[main node, scale=0.8] (8) at (4,0.5) {8};
  \node[main node, scale=0.8] (9) at (3,1.5) {9};
  \node[main node, scale=0.8] (10) at (5,1.5) {10};

  \path[every node/.style={font=\sffamily\small}]
    (1) edge [right] node[left] {} (2)
        edge [right] node[left] {} (3)
        edge [right] node[left] {} (4)
        edge [right] node[left] {} (5)
    (2) edge node {} (4)
    (3) edge node [right] {} (5)
        edge node [right] {} (6)
    (4) edge node [right] {} (5)
    (6) edge node [right] {} (7)
    (7) edge node [right] {} (8)
    (8) edge node [right] {} (9)
        edge node [right] {} (10)
    (9) edge node [right] {} (10)
     ;
\end{tikzpicture}
            }} & 
\subfigure[Centrality calculation.]{
\resizebox{0.45\linewidth}{!}{
                \begin{tikzpicture}[-,>=stealth',shorten >=1pt,auto,node distance=3cm,scale=0.6,
                        thick,main node/.style={circle,draw}]
    
      \node[main node, scale=0.8, draw=red!60] (1) at (0,0) {1};
      \node[main node, scale=0.8, draw=red!60] (2) at (-1,1) {2};
      \node[main node, scale=0.8, draw=red!60] (3) at (1,-1) {3};
      \node[main node, scale=0.8, fill=red!30] (4) at (0.5,1.5) {4};
      \node[main node, scale=0.8, draw=red!60] (5) at (1.5,0.5) {5};
      \node[main node, scale=0.8, draw=red!60] (6) at (2.5,-1.5) {6};
      \node[main node, scale=0.8] (7) at (4,-1) {7};
      \node[main node, scale=0.8] (8) at (4,0.5) {8};
      \node[main node, scale=0.8] (9) at (3,1.5) {9};
      \node[main node, scale=0.8] (10) at (5,1.5) {10};
    
      \path[every node/.style={font=\sffamily\small}]
        (1) edge [right] node[left] {} (2)
            edge [right, red!60] node[left] {} (3)
            edge [right, red!60] node[left] {} (4)
            edge [right] node[left] {} (5)
        (2) edge[red!60] node {} (4)
        (3) edge node [right] {} (5)
            edge[red!60] node [right] {} (6)
        (4) edge[red!60] node [right] {} (5)
        (6) edge node [right] {} (7)
        (7) edge node [right] {} (8)
        (8) edge node [right] {} (9)
            edge node [right] {} (10)
        (9) edge node [right] {} (10)
         ;
    \end{tikzpicture}
                }} \\
\multicolumn{2}{c}{
\subfigure[Aggregation points selection.]{
\resizebox{0.45\linewidth}{!}{
                \begin{tikzpicture}[-,>=stealth',shorten >=1pt,auto,node distance=3cm,scale=0.6,
                        thick,main node/.style={circle,draw}]
    
      \node[main node, scale=0.8, fill=gray!30] (1) at (0,0) {1};
      \node[main node, scale=0.8, fill=gray!30] (2) at (-1,1) {2};
      \node[main node, scale=0.8, fill=gray!30] (3) at (1,-1) {3};
      \node[main node, scale=0.8, fill=green!30] (4) at (0.5,1.5) {4};
      \node[main node, scale=0.8, fill=gray!30] (5) at (1.5,0.5) {5};
      \node[main node, scale=0.8, fill=green!30] (6) at (2.5,-1.5) {6};
      \node[main node, scale=0.8, fill=gray!30] (7) at (4,-1) {7};
      \node[main node, scale=0.8, fill=gray!30] (8) at (4,0.5) {8};
      \node[main node, scale=0.8, fill=green!30] (9) at (3,1.5) {9};
      \node[main node, scale=0.8, fill=gray!30] (10) at (5,1.5) {10};
    
      \path[every node/.style={font=\sffamily\small}]
        (1) edge [right] node[left] {} (2)
            edge [right] node[left] {} (3)
            edge [right] node[left] {} (4)
            edge [right] node[left] {} (5)
        (2) edge node {} (4)
        (3) edge node [right] {} (5)
            edge node [right] {} (6)
        (4) edge node [right] {} (5)
        (6) edge node [right] {} (7)
        (7) edge node [right] {} (8)
        (8) edge node [right] {} (9)
            edge node [right] {} (10)
        (9) edge node [right] {} (10)
         ;
    \end{tikzpicture}
                
        }}
}
\end{tabular}
    \end{center}
    \caption{Aggregation points selection based on $k$-closeness centrality.}
    \label{fig:sketch}
\end{figure}

Figure \ref{fig:sketch} illustrates the steps to solve our problem for a minimum one-hop dominating set, with closeness centrality set to $k = 3$. 
As depicted in Figure \ref{fig:sketch}(b), the closeness centrality of vertex 4, which is equal to $0.125$ (the highest value), only considers the vertices highlighted in red. 
An aggregation point collects and aggregates data from its $d$-hop neighbors. 
Here, we select three vertices (4, 6, and 9) as aggregation points, but we can find other solutions by adjusting the parameters. In addition, more than one $v_i$ can maximize $C_k$. We deal with this by choosing the first one.

The data aggregation process may consider multi-hop communication. 
The proposed solution starts with local communication between each pair of vehicles to model the network and extracts information from the graph.
The execution of the algorithm must precede each data delivery phase. 
This approach leads to the following main advantages:
(i) The solution addresses single and multi-hop (with a different number of hops) communication to maximize cellular bandwidth saving;
(ii) we can adjust the calculation of centrality value according to traffic conditions and application requirements;
(iii) the heuristic works separately at each base station, favoring scalability and decentralized management. In the case where all vehicles are isolated, each one will transmit its data individually, resulting in an upload cost equivalent to the traditional approach; and
(iv) vehicles operate asynchronously. Thus, scenarios with a high density of vehicles do not offer synchronization overhead.

The centrality-based algorithm does not guarantee the optimal solution in all cases but provides approximate solutions in polynomial time. 
In addition to the cost of $\mathcal{O}(n{\lambda}^k + n\log n)$ to discover vehicles with high centrality values, there is a cost $\mathcal{O}(n{\lambda}^d)$ to find neighbors up to the \textit{d-th} level. 
The algorithm's time complexity to solve the \textit{minimum d-hop dominating set} for a $\lambda$-regular graph is $\mathcal{O}(n(\lambda^d + \lambda^k + \log n))$.

Some vehicles chosen as aggregation points may be traveling in opposite directions as the other vehicles in the dominating set, resulting in a short link period with the aggregation points leaving the group of vehicles. 
Mobility information can complement the centrality measure to increase communication reliability and cluster stability.
We will calculate the centrality measure using Equation~\ref{equa:close2} in a new graph with edge constraints, where only paths formed by vehicles with the exact direction will compose the network. This restriction ensures that each community of nodes in the network will consist only of vehicles traveling in the same direction.
Let $w_i = P_{2} - P_{1}$ be the displacement vector of the vehicle $v_i$, in which $P_{2}$ is the coordinates of the current position of $v_i$ and $P_{1}$ is the coordinates of the last position. The vector $w_i$ indicates the distance and direction traveled by $v_i$ from $P_1$ to $P_2$.
The angle of rotation ($\theta$) between $w_i$ and $w_j$ is given by:
\begin{equation}
    \theta = \arccos \frac{w_i \boldsymbol{\cdot} w_j}{|w_i||w_j|},    
\end{equation}

where $w_i \boldsymbol{\cdot} w_j$ is the dot product of two vectors and $|w_i|$ and $|w_j|$ is the magnitude of $w_i$ and $w_j$, respectively. The vehicles $v_i$ and $v_j$ have different directions if $\theta$ exceeds a threshold. The threshold can be between 0º and 180º because the method returns the smallest angle between the vectors. Then, 180º corresponds to the opposite direction. If $\theta \leq$ 45º, the two vehicles have the same direction (a value typically used in the literature~\cite{tal2016}).
In this manner, two vehicles, $v_i$ and $v_j$, will be neighbors if they are in the same transmission range and have the exact direction. We can calculate vehicle directions by traversing the adjacency list at $\mathcal{O}(m)$.

\subsection{Method for Parameter Selection}\label{sec:optimization}

Finding an optimal choice of parameters is an essential design issue. In numerical analysis, the researchers use widely \textbf{Nelder-Mead simplex algorithm} to solve parameter estimation problems~\cite{nelder1965}. It is a usual search method to optimize an objective function in a p-dimensional space based on the iterative update of a simplex with ($p + 1$) points. The simplex associates each point with a function value and sorts, in linear time, according to these values. At each iteration, we use reflection, expansion, or contraction to replace the worst point with a better one. The updating executes in $\mathcal{O}(p)$ operations. We repeat this process until to reach one of the stopping criteria, e.g., the maximum number of iterations.
We use the Nelder-Mead method to find the values of the parameters that maximize the aggregation rate function. The parameter $d$ sets the number of communication hops, whose maximum value is $max_{i,j} \ \mbox{dist}(v_i, v_j) \ \forall i, j = \{1, ..., n\}$, i.e., the network diameter. As we will see later, there is a trade-off between the aggregation rate and the end-to-end delay. Therefore, we must use typically small values for the number of hops to save bandwidth and reduce latency. Usually, the maximum number of hops between a source node and a destination is ten on average~\cite{sommer2007}. In this manner, the centrality-based algorithm uses this value as an upper boundary for the number of hops. 

\section{Evaluation and results}\label{sec:experiments}

This section describes the performance evaluation of the \textbf{Centrality-based algorithm}. 
We used a realistic representation of vehicular mobility to model D2D communication.
The evaluations consider simulation experiments to analyze different performance metrics. 

\subsection{Scenarios}
We consider two main scenarios:
(i) \textit{Centralized}: we use the coverage of only one cellular cell with one server processing and determine the aggregation points in our application. 
In this case, we compare our algorithm's performance against the RB algorithm (proposed by Stanica et al. ~\cite{Stanica2013}), the optimal solution, and the traditional upload.
(ii) \textit{Decentralized}: we use the coverage of four cellular cells with decentralized server processing and determine the aggregation points in our application. We use the RB algorithm as the baseline. 

Given its high simulation cost, we use a large-scale urban region with a simplified transmission model in the first scenario. The second is a more realistic but smaller scenario, considering a multi-cell infrastructure and a detailed LTE model.
In both scenarios, we use a dataset based on realistic traffic demand in Cologne, Germany.

This dataset is available through the TAPASCologne project \citep{Uppoor2014}, an Institute of Transportation Systems (ITS-DLR) initiative at the German Aerospace Center. 
We perform all experiments in a computer with a Ubuntu 16.04 operating system, Intel Core i5-7200U @2.50 GHz processor, and \unit[8]{GB} of RAM.

We assume sensory data collected in close space-time proximity has a significant correlation, such as temperature measurements.
Thus, we aggregate data with a simple local aggregation model based on the average to avoid requiring each vehicle's data.
Each aggregation point aggregates data from its neighbors (including its data) and transmits a single packet summarizing the collected data. 
The packets have a fixed size of $\epsilon =$ 120 bytes. 
The number of packets ($p$) changes over time according to the size of the dominating set $S$, resulting in a total data volume of $p \epsilon$ bytes uploaded in each data delivery period (\unit[10]{s}).
    
We evaluate the \textbf{Centrality-based algorithm} with different number of hops $d = \{1, 2, ..., 6\}$ and closeness depth $k = \{1, 2, ..., 6\}$. We fixed the number of time slots in the RB algorithm $T = 256$, as defined in~\cite{Stanica2013}.
We perform ten simulation runs for each parameter value and calculate the average. 
The acceptable minimum number of simulations is
%
    $\text{rounds} = (100\, z\, \sigma/\mu)^2$~\citep{jain1990art},
%
where $z = 1.96$ is a critical value for a 95\% confidence level, and $\mu$ is the value required for a 5\% error margin. 
We obtain the sample standard deviation ($\sigma$) from the five worst cases of 10 preliminary independent simulations (with a unique seed for each one).

We apply two statistical tests to analyze the results: 
(i) \textit{Shapiro-Wilk test}~\cite{shapiro} to check the normality of data distribution. The null hypothesis for the Shapiro-Wilk test is that some populations have the samples normally distributed. In contrast, if \textit{p-value} $ < 0.05$ for a significance level ($\alpha$) of $0.05$, we reject the null hypothesis (not normal samples).  
According to the test's result for normality, the samples do not follow a normal distribution, so we evaluated them with (ii) the \textit{Wilcoxon signed-rank test}~\cite{wilcox} with a 95\% confidence level. It is a non-parametric paired test used with not normally distributed data. We used the paired test to determine whether the difference between the two dependent samples follows a symmetric distribution around zero (i.e., samples selected from populations with the same distribution).

\subsection{Centralized scenario}\label{sec:centralized}

In our centralized scenario evaluations, we use the complete Cologne's trace\footnote{http://kolntrace.project.citi-lab.fr/}, which covers \unit[400]{km$^2$} through 24 hours with a granularity of one second. 
We use more than 700,000 individual car trips and mobility traces to model vehicle-to-vehicle communication (e.g., based on the IEEE 802.11p standard or D2D communication).
The connections between vehicles follow the Unit Disk Graph (UDG) model, a simple and popular model for wireless communication networks. In this model, two nodes communicate if the distance between them is most $r$, where $r$ is the transmission radius equal for all nodes. We assume all vehicles are homogeneous and equipped with omnidirectional antennas with a transmission range of \unit[100]{meters}.
We can not guarantee 100\% coverage for all vehicles in a real scenario, but we consider only a cellular base station providing ubiquitous coverage to all vehicles. With this simple scenario, we can simulate the aggregation points selection in a centralized manner, in which each vehicle has a cellular network interface to upload its sensory data. We fixed the angle threshold as 45º, defined in ~\cite{tal2016}. So, when the rotation angle between two vehicles is more than 45º, we assumed these vehicles are not on the same road and have different directions.

The metrics used for the performance evaluation in this scenario are
(i) \textit{Aggregation rate}: the ratio between data volume after aggregation and data volume before aggregation. We used this metric to evaluate the gain of the offloading scheme;
(ii) \textit{Upload cost}: the amount of sensory data uploaded for each data delivery. The higher the aggregation rate, the better the upload cost reduction; 
(iii) \textit{Computational cost:} the total number of edges examined to estimate the $k$-closeness centrality. It shows the impact of the cutoff ($k$) parameter value on the cost of running the algorithm;
(iv) \textit{number of edges:} the total number of connections between the vehicles in the network. We use this measure to analyze the impact of vehicle directions on network connectivity;
(v) \textit{number of reelections:} the average number of vehicle reelections as an aggregation point. It measures the efficiency of selecting aggregation points where a low number of reelections suggests that cluster structures are unstable to minor variations in the network structure; (vi) \textit{number of notification messages:} the average number of notification messages sent from the base station to the aggregation points. We used this metric to evaluate the cost of forming the clusters during the aggregation points selection process; and (vii) \textit{number of routing updates:} average number of updates in the routing table of vehicles. It measures the cost of maintaining the cluster after updating the aggregation point.

\begin{figure}[htb]
    \begin{center}
        \subfigure[CDF of the aggregation rate.]{
        \includegraphics[width=\linewidth]{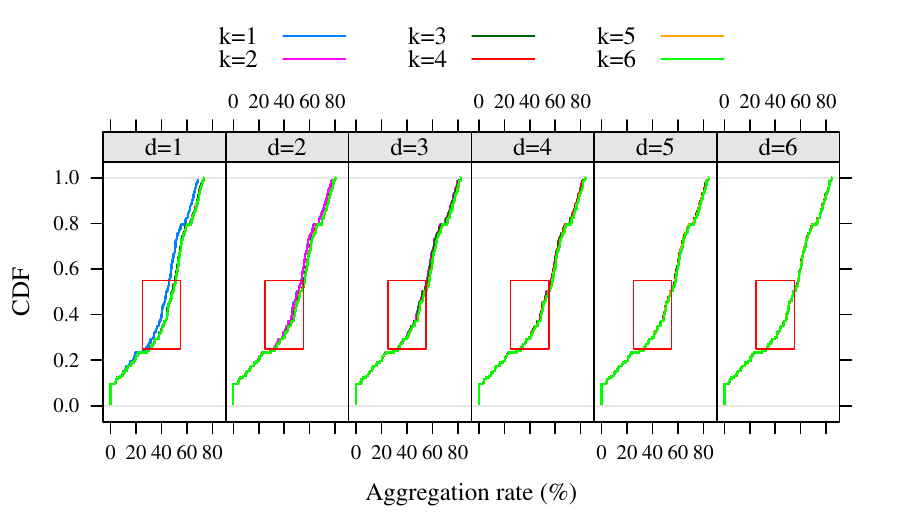}
        }
        \subfigure[Zoom of the CDF of the aggregation rate.]{
        \includegraphics[width=\linewidth]{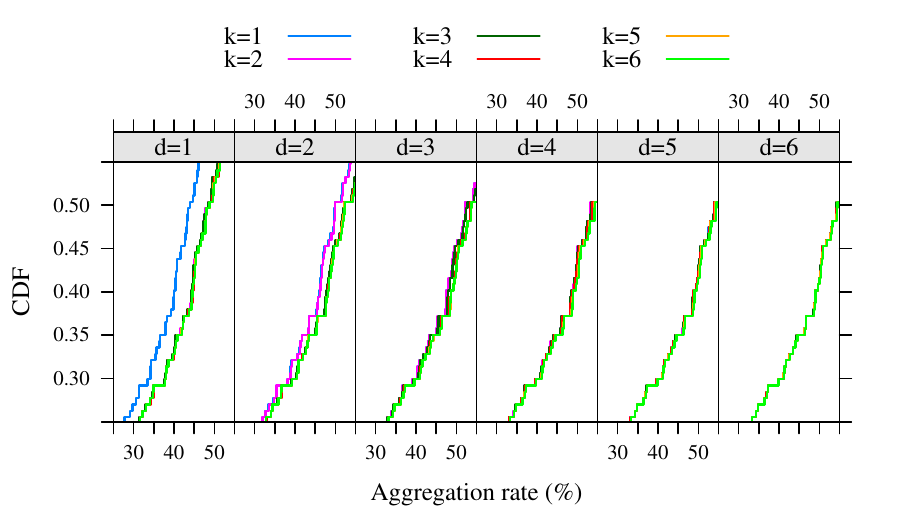}
        }
    \end{center}
    \caption{Cumulative distribution function of the aggregation rate of the centrality-based algorithm.}
    \label{fig:cdf}
\end{figure}

Figure \ref{fig:cdf} presents the aggregation rate's Cumulative Distribution Function (CDF) for the centrality-based algorithm with different parameter values.
The CDF function shows that the probability of aggregation rate is less than or equal to a specific value. 
We observe that the number of hops ($d$) and the maximum path length ($k$) affect the aggregation rate of the data offloaded by vehicles.
When $k$ is significantly tiny (i.e., $k \ll$ maximum path size), the values obtained for the centrality measure are underestimated compared to the actual values. This underestimation occurs because the calculation of closeness centrality considers only partial information about the network topology.
The results suggest that the aggregation rate is better when $k > d$ since the nearest neighbors hardly provide more information to estimate a vertex's centrality measure.
Therefore, we use $k = 4$ to analyze single ($d = 1$) and multi-hop ($d = 3$) communications. This value provides more accurate information to calculate the centrality of a vehicle considering its $k$-neighborhood.

\begin{figure}[htb]
    \begin{center}
        \includegraphics[width=0.92\linewidth]{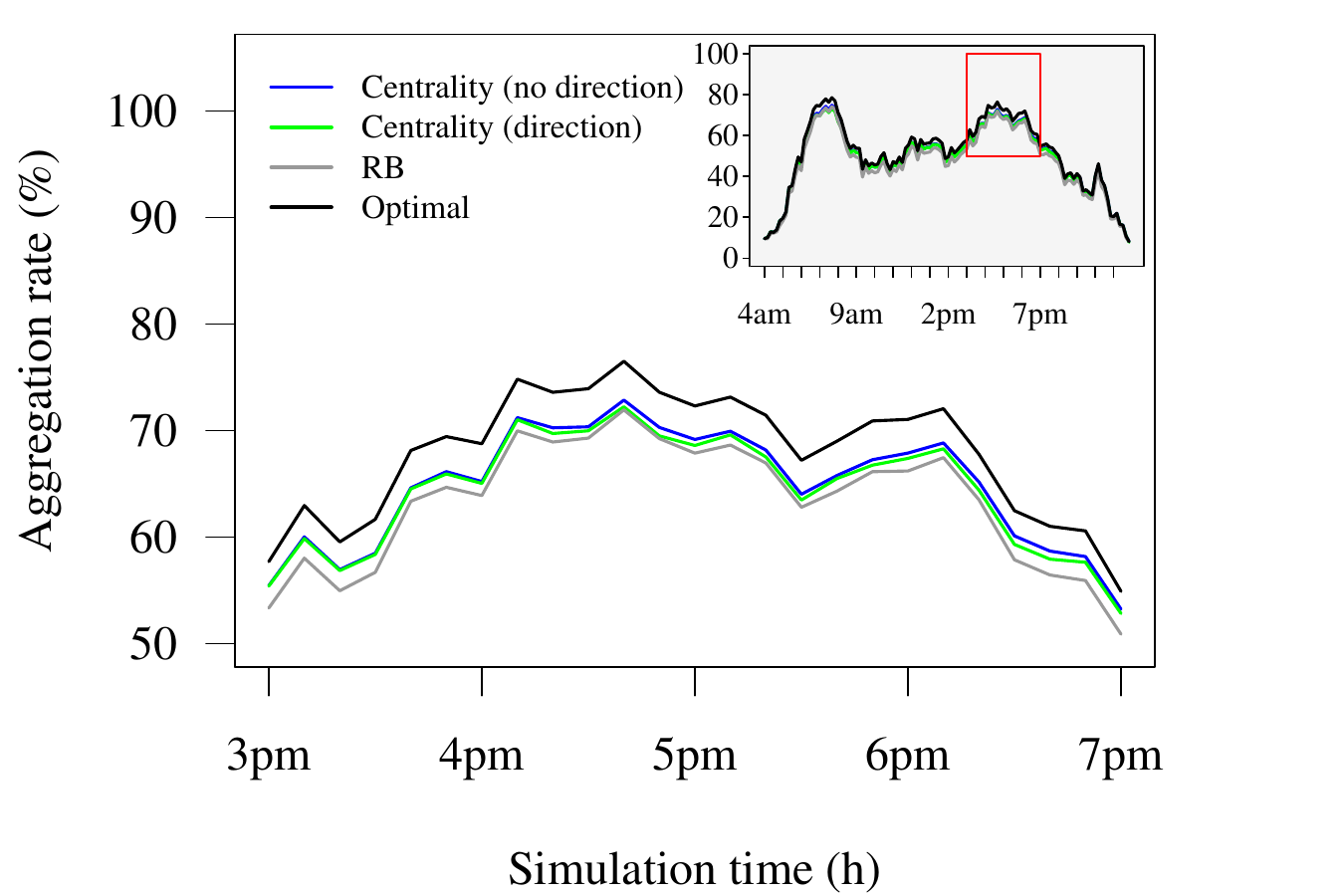}
        \end{center}
        \caption{Aggregation rate obtained using a centrality-based algorithm compared to the RB and the optimal solutions (single-hop communication).}
        \label{fig:single}
\end{figure}

Figure \ref{fig:single} presents the aggregation rate results of the optimal solution, RB algorithm, and the Centrality-based algorithm in a single-hop communication. In addition, we also run the centrality-based algorithm without vehicle directions information.
We can observe that both Centrality-based algorithms are better than RB. Our algorithm minimizes the number of hops between the aggregation points and their neighbors whenever it is possible to cover all vehicles with a small number of aggregation points. The optimal solution gives the minor aggregation points set a \textit{minimum $d$-hop dominating set}. The RB algorithm constantly approximates the optimal solution when the number of transmission slots tends to infinity. However, in a real scenario, the number of transmission slots is limited, resulting in sub-optimal performance of the RB algorithm~\cite{6814727}. This limitation is because a vehicle in the \texttt{contender} state can receive more than one reservation message transmitted at the same transmission slot, which can cause collisions. In addition, this vehicle might not decode some messages and enter the \texttt{dominator} state.
The results also show a decrease in the aggregation rate when we consider only vehicles in the same direction during offloading. This behavior occurs because an aggregation point cannot collect data from vehicles in different directions. In the worst case, we will need more aggregation points to ensure complete coverage of vehicles.

Next, we apply the One-Tailed version of the Wilcoxon signed-rank test to compare the aggregation rates observed in the algorithms based on centrality and reservation, which reported $p$-values of 2.2e-16 (less than $0.05$). We can reject the null hypothesis in favor of the alternative one.
Thus, the test indicates that the centrality-based algorithm has a better aggregation rate than the RB algorithm. Furthermore, the reported results suggest that disregarding vehicle directions does not invalidate the solution regarding aggregation rate. Therefore, the centrality-based algorithm can run even when vehicle directions information is not available. As we will see in the next section, clusters are more stable than those formed by the RB. In this case, the vehicles selected as aggregation points often remain in the cluster and are reelected.
We can also see that the \textbf{Centrality-based algorithms} find an optimal solution for some instances.

        

\begin{figure}[t]
    \begin{center}
        \includegraphics[width=.92\linewidth]{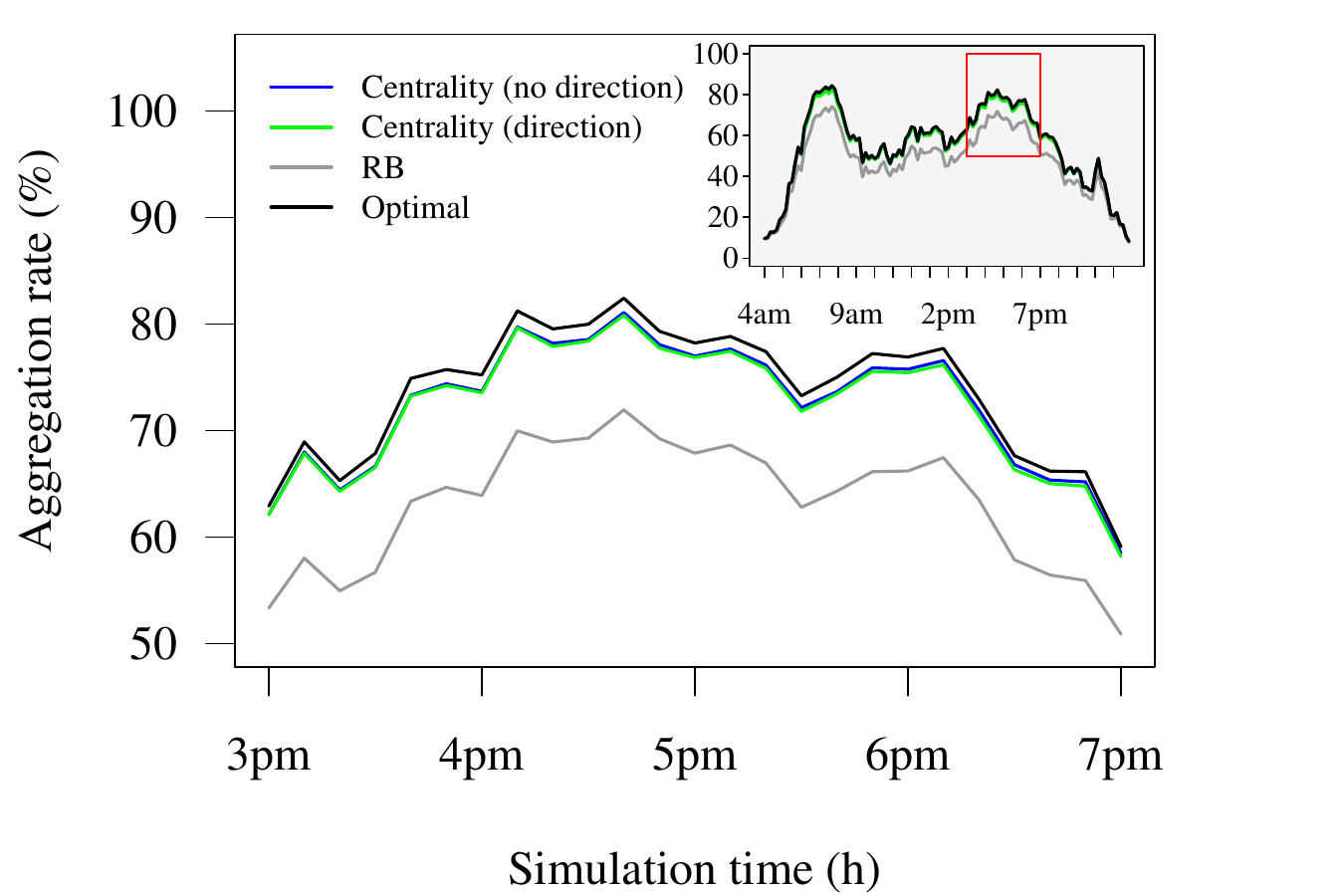}
        \end{center}
        \caption{Aggregation rate obtained using a centrality-based algorithm compared to the RB and optimal solutions (multi-hop communication).}
        \label{fig:multi}
\end{figure}

We can explore growing neighborhoods by increasing the number of hops. 
However, the network's diameter limits the benefit of multi-hop communication.
Figure \ref{fig:multi} presents the aggregation rate results for $d = 3$.
Again, the centrality (no direction) approach performed better than the centrality with direction. The direction constraint eliminates edges in the graph and causes a slight reduction in the aggregation rate. In this scenario, there was an average reduction of 2.16\% in the number of edges. This behavior suggests that a few vehicles move in different directions, resulting in an average difference of 0.17\%.
Compared with single-hop communication, the Centrality-based algorithm without direction improves the aggregation rate by 8.27\% at peak hours and has nearly optimal performance.
We keep the RB aggregation rate in this figure, in which the centrality without direction, when compared with the RB solution, achieves an aggregation rate increase of up to 10.45\%.
The RB algorithm's minimum distance between aggregation points is two hops (best case). However, the domination number of the network will limit the number of aggregation points selected. The multi-hop communication can increase the distance between aggregation points, where fewer aggregation points are needed to cover vehicles. Moreover, the RB algorithm does not consider topological information from the network and selects the aggregation points according to randomly chosen time slots. As a result, the sets of selected aggregation points tend to be larger than those of the centrality-based algorithm.

\begin{figure}[t]
    \begin{center}
        \subfigure[Computational cost of parameter optimization.]{        \includegraphics[width=.8\linewidth]{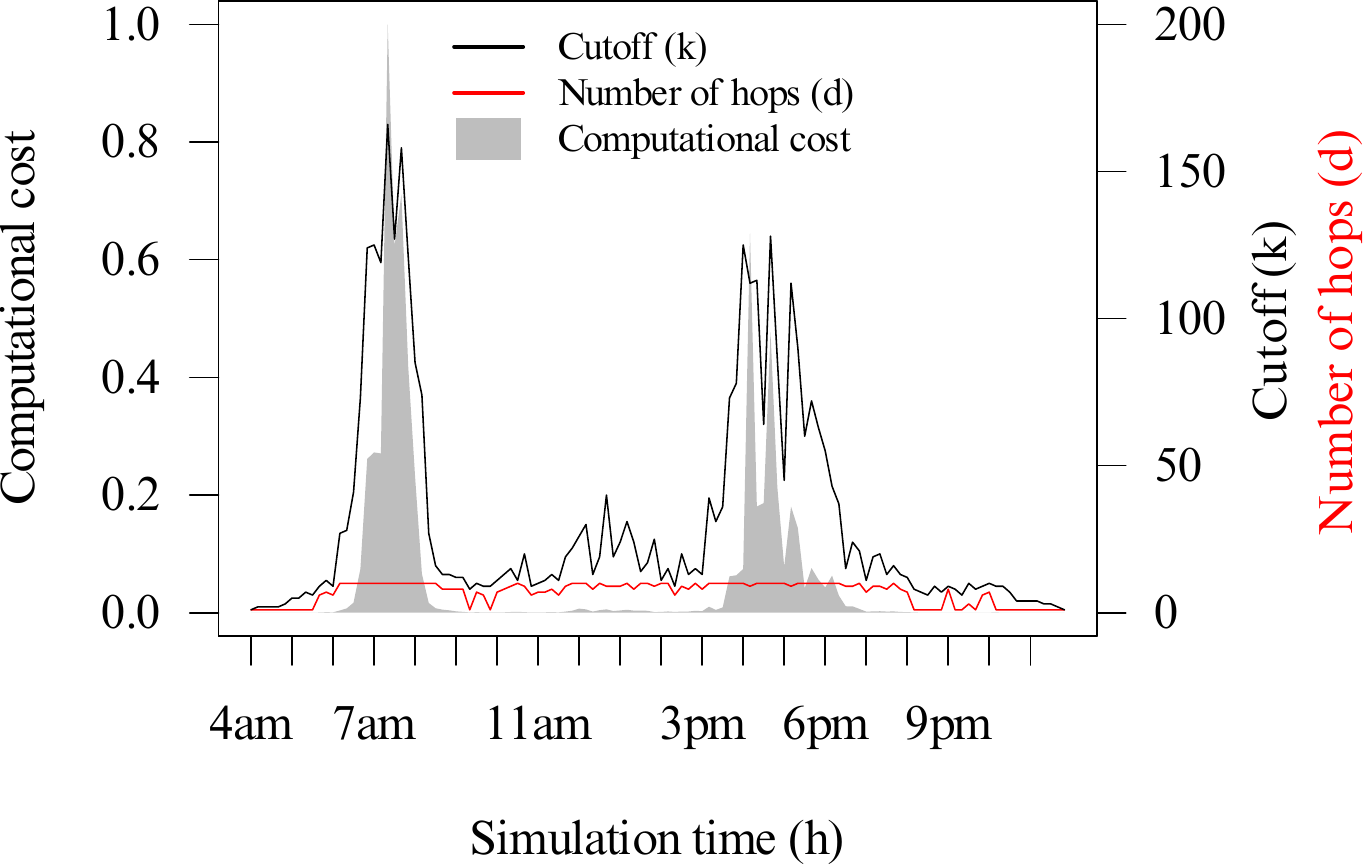}
            \label{fig:optimal_k}
        }
        \\
        \subfigure[Aggregation rate of parameter optimization.]{            \includegraphics[width=.8\linewidth]{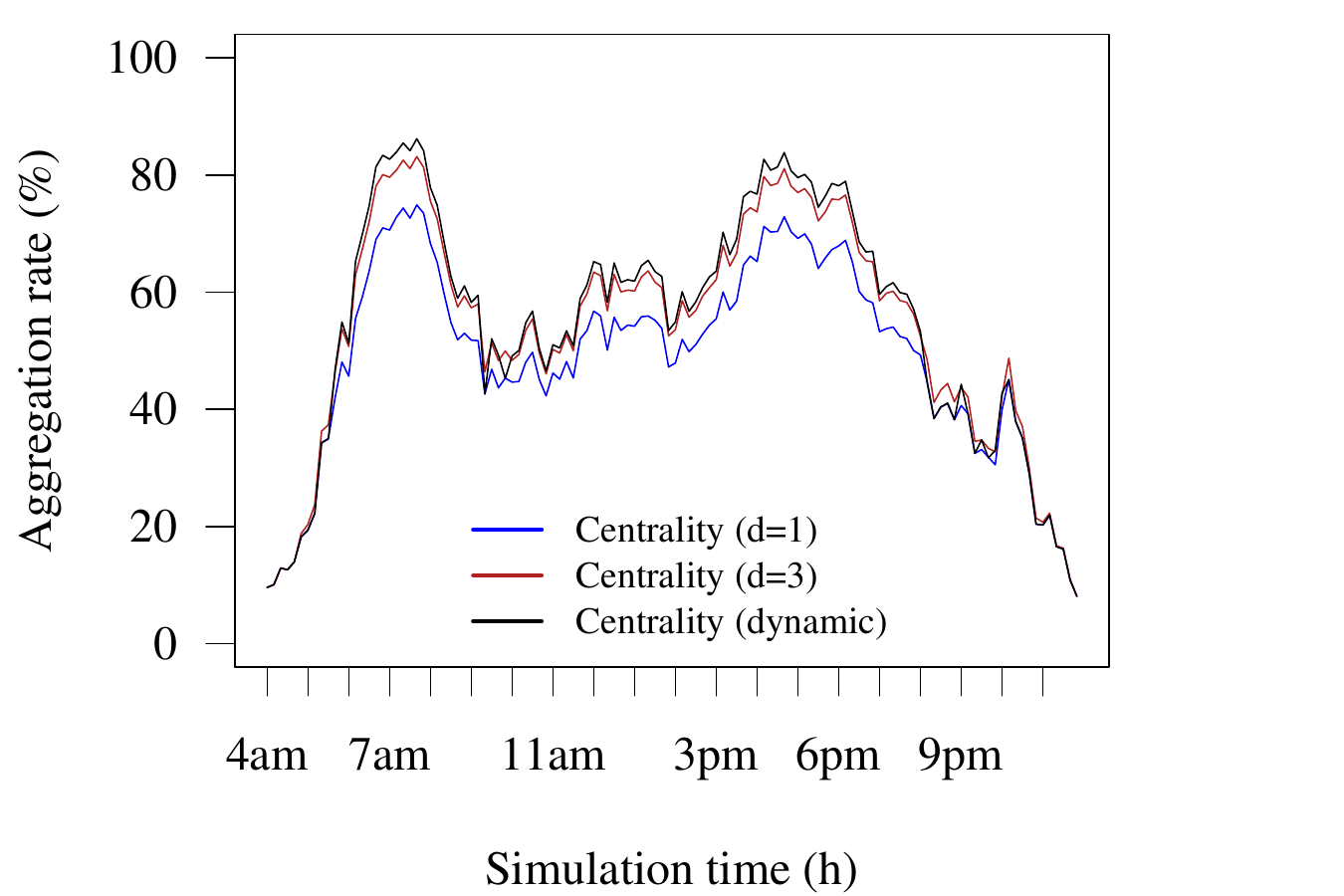}
            \label{fig:optimal_ag_rate}
        }
    \end{center}
    \caption{Computational cost and aggregation rate of parameter optimization (Nelder-Mead method).}
\end{figure}

Figure~\ref{fig:optimal_k} presents the optimization results for the Centrality-based algorithm using the Nelder-Mead method (see Section~\ref{sec:optimization}) with a limit of 500 iterations. Most parameter values $k$ are between 6 and 28, with extreme values occurring during peak hours. We analyze the computational cost based on the number of edges examined during the closeness centrality estimation. A large $k$ leads to a high cost in dense networks with an expressive number of connections, which may need to improve in applications with strict latency requirements. On average, the centrality-based algorithm used six hops for communication, with an increase during peak hours, when the network becomes densely connected. Figure~\ref{fig:optimal_ag_rate} shows the aggregation rate after the parameters optimization. This procedure brought a significant improvement to the Centrality-based algorithm. First, however, we must recognize the high cost entirely. This figure shows that the Centrality-based algorithm ($d = 3$) achieved an aggregation rate close to the one with optimized parameters, suggesting we can obtain a satisfactory aggregation rate without an exhaustive search.

\begin{figure}[t]
    \begin{center}
    \subfigure[Cost of upload.]{
    \includegraphics[width=.8\linewidth]{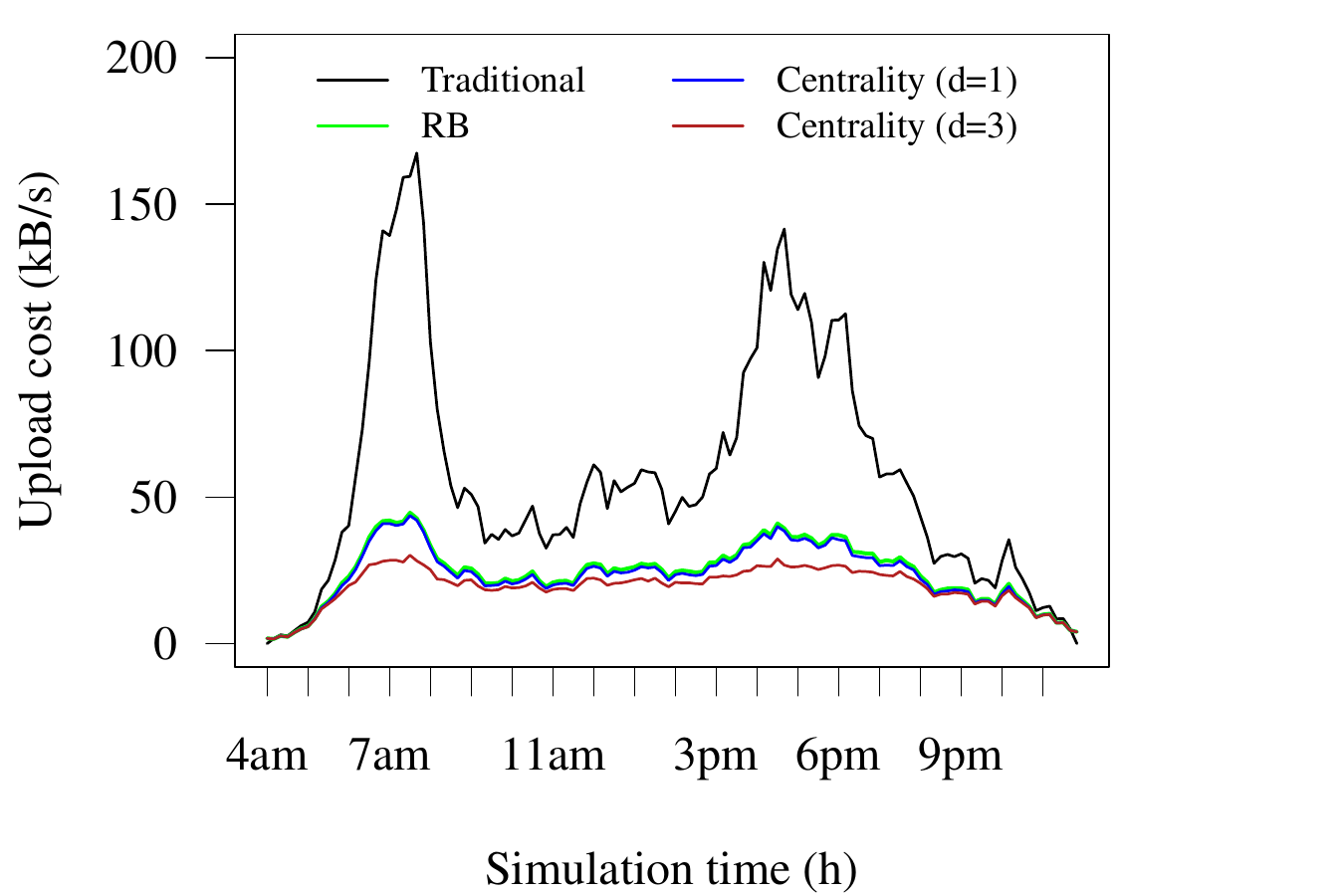}
    }
    \\
    \subfigure[Number of vehicles.]{
    \includegraphics[width=.8\linewidth]{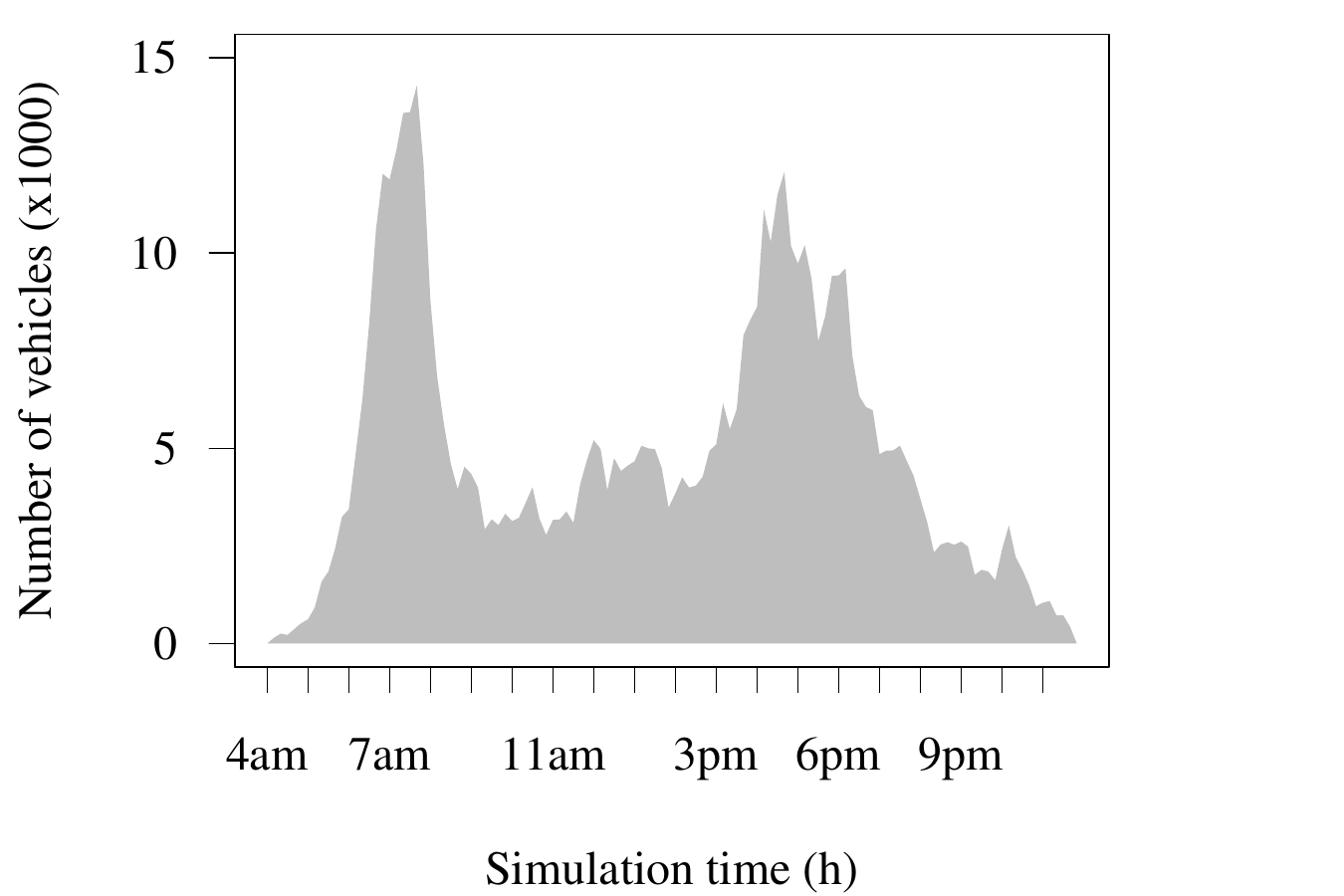}
    }
    \end{center}
    \caption{cost of upload in the traditional upload approach.}
    \label{fig:no_off}
\end{figure}

Figure \ref{fig:no_off} shows the traditional approach's upload cost (no offloading scheme), where each vehicle uploads its data.
The upload cost for each data delivery period is the sum of all data sent by each vehicle in the scenario. In this case, the upload cost is directly proportional to the number of vehicles.
To traditional upload, the cost is \unit[167.50]{kB/s} at peak hours, and after performing offloading with the \textbf{Centrality-based algorithm} (no direction), \unit[43.66]{kB/s} -- a reduction of 73.93\%.
The upload cost is similar to the RB algorithm, \unit[44.69]{kB/s} -- a reduction of 73.31\%. However, we obtained an essential improvement over the RB algorithm.
Upload's cost decreased from \unit[43.66]{kB/s} to \unit[30.16]{kB/s} after $d = 3$ hops, which represents a reduction of 81.99\% and 30.92\% in the upload cost compared with the traditional approach and the RB algorithm, respectively.

In all cases, the upload cost is higher at peak hours, where the results show the best aggregation rate and cost reduction.
The throughput in the D2D network is $p\epsilon/10$ bytes per second, where $p$ is the number of generated packets, and 10 is the collect interval.
In both approaches, the bandwidth consumption is higher when the number of nodes increases. However, the RB Algorithm must transmit the reservation messages, where all vehicles can choose the same time slot. Therefore, in the worst case, we send $p = n$ packets on the D2D network, which occurs with a probability $T^{1-n}$. In contrast, there is an overhead in the Centrality-based algorithm to relay sensing data. Therefore, the generated traffic will be higher when the multi-hop communication is enabled, resulting in $p = n^d$ packets. With this increment, we save the cellular uplink bandwidth, with an aggregation rate of 10.45\% compared to the RB algorithm. In the next section, we will look at the decentralized scenario and the impact of multi-hop communication.

\begin{figure*}[htb]
    \begin{center}
        \subfigure[Aggregation rate for single-hop communication.]{        \includegraphics[width=.45\linewidth]{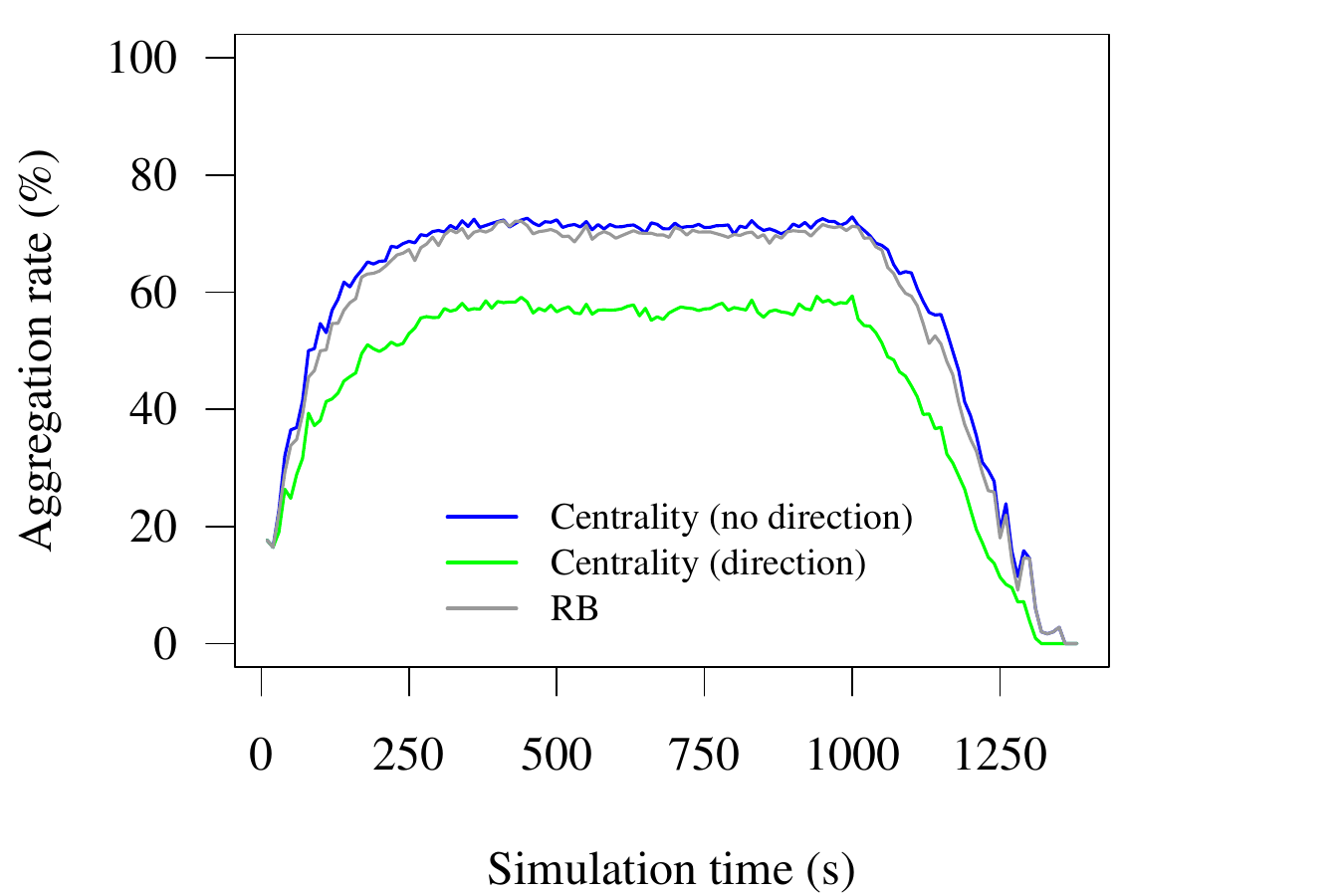}
            \label{fig:dir1}
        }
        \subfigure[Number of edges in the network.]{            \includegraphics[width=.45\linewidth]{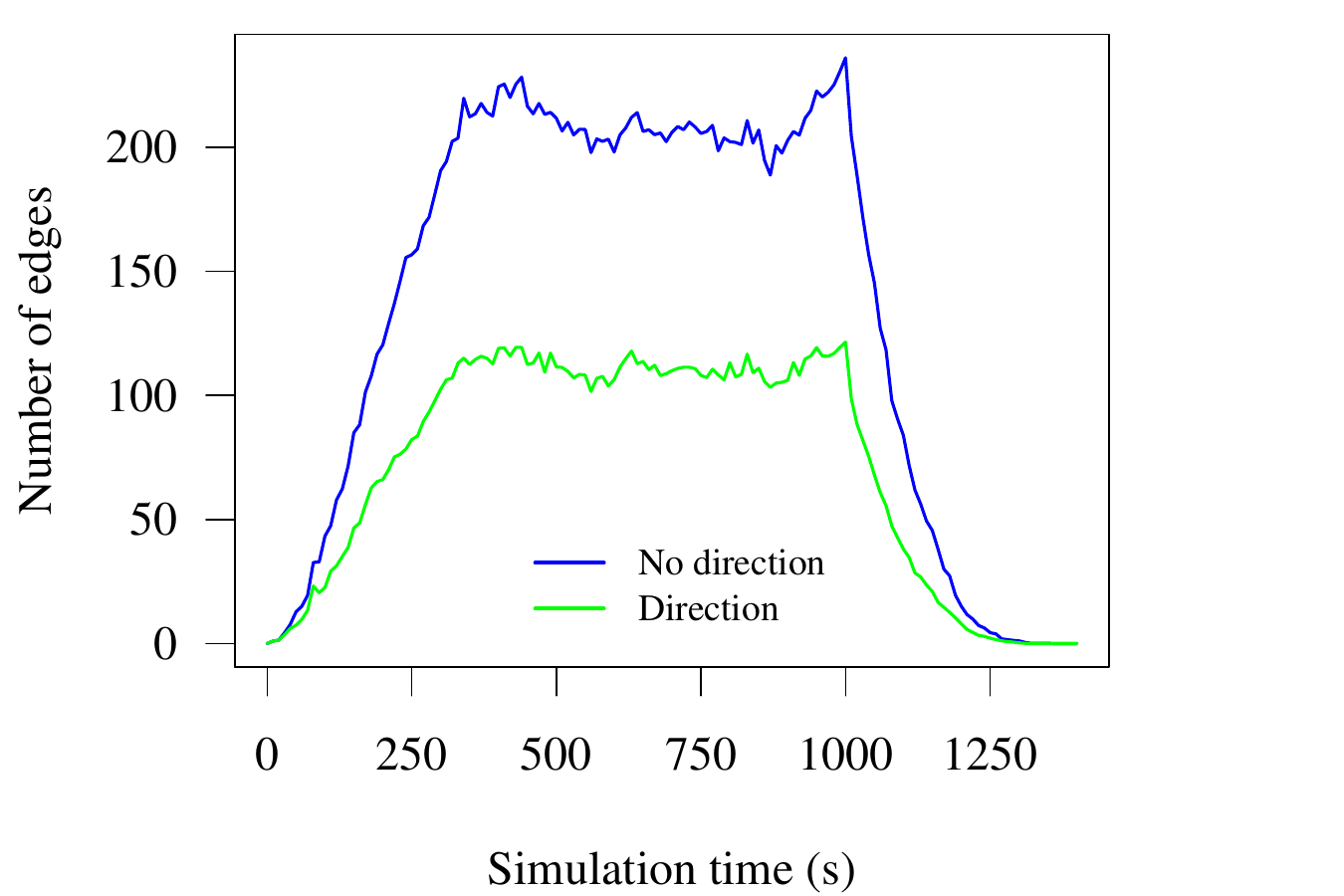}
            \label{fig:dir2}
        }
    \end{center}
    \caption{Evaluation of the two-way roadway scenario.}
\end{figure*}

Although the dataset presents a real scenario with different road topologies, we need to evaluate the performance of the proposed algorithm in a two-way roadway scenario.
We created a square area with dimension $\unit[15]{km} \times \unit[15]{km}$, with 1.500 vehicles for 1.400 seconds. This scenario consists of a two-way roadway, randomly generating routes by \textbf{SUMO random trips tool}. We perform the aggregation points selection using Centrality-based and RB algorithms. For this experiment, we used the same parameters as in the previous one.
Figure \ref{fig:dir1} presents the aggregation rate analysis.
When we do not consider the direction of the vehicles, the Centrality-based algorithm without direction performs better than the RB algorithm. However, if we include the direction of the vehicles as a constraint for the Centrality-based algorithm, then the aggregation rate will be penalized. Figure \ref{fig:dir2} shows the number of edges in the network during the simulation. We can observe that the directions impose severe constraints on the network structure, resulting in decreased network connectivity. Different of Cologne scenario, this scenario has approximately half of the vehicles traveling in different directions, resulting in many edge removals. In an exploratory study, we identify that the average percentage of removed edges in the Cologne scenario was 2.16\% and in the two-way roadway scenario, was 48.22\%. Consequently, many aggregation points are needed to ensure complete network coverage, leading to an average reduction of 13.37\% in the aggregation rate. Although the aggregation rate is lower, we see an increase in network stability. This behavior is because a vehicle tends to remain longer as an aggregation point when we calculate vehicle directions. On average, an aggregation point was reelected $3.66 \pm 0.27$ times in the Centrality-based algorithm with direction. When we do not consider the vehicles' directions, the average number of reelections was $0.87 \pm 0.07$ in the Centrality-based algorithm without direction and $0.66 \pm 0.05$ in the RB algorithm.

\begin{figure}[htb]
    \begin{center}
        \subfigure[Number of notification messages.]{        \includegraphics[width=.95\linewidth]{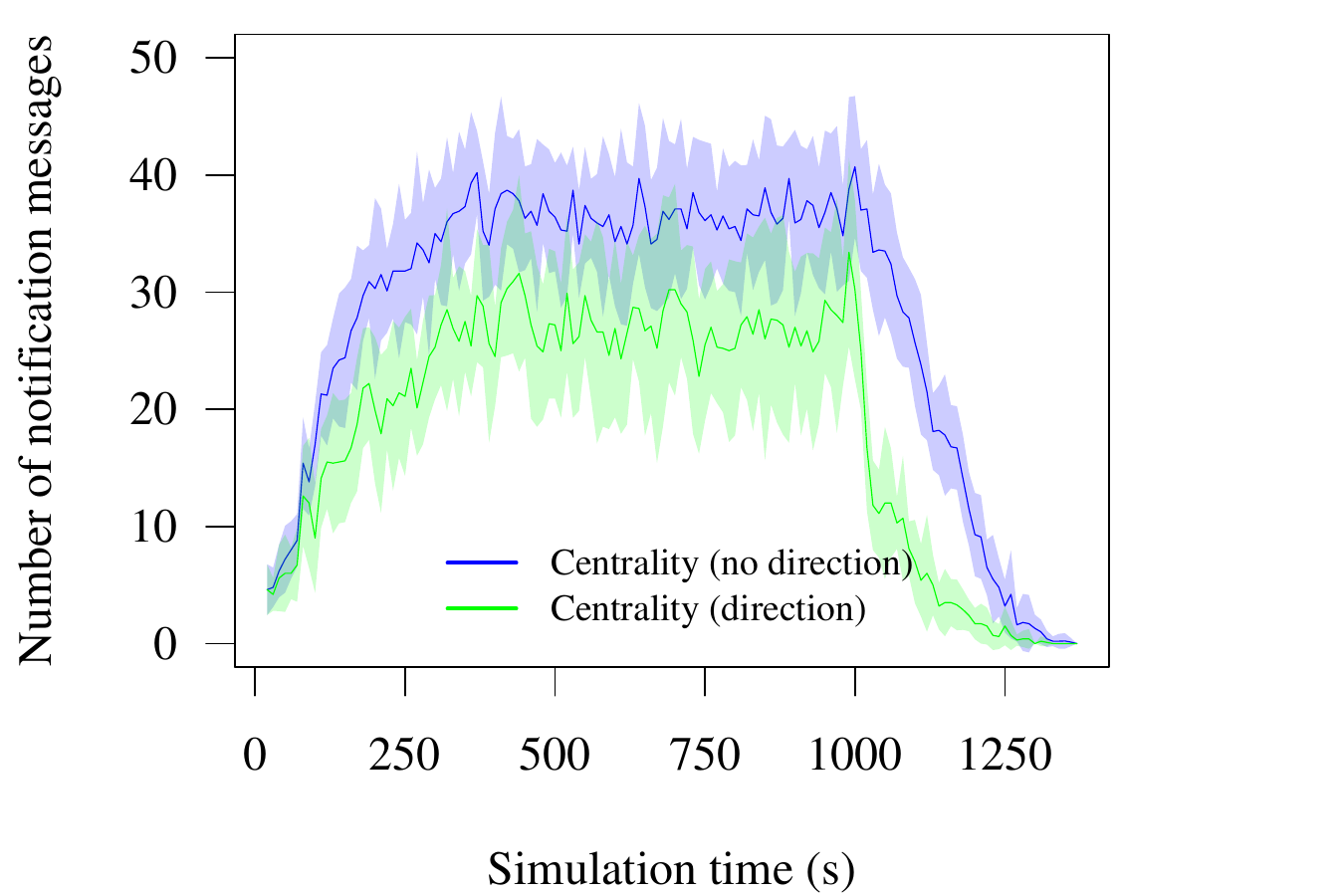}
            \label{fig:cost1}
        }
        \\
        \subfigure[Number of Routing Updates.]{            \includegraphics[width=.95\linewidth]{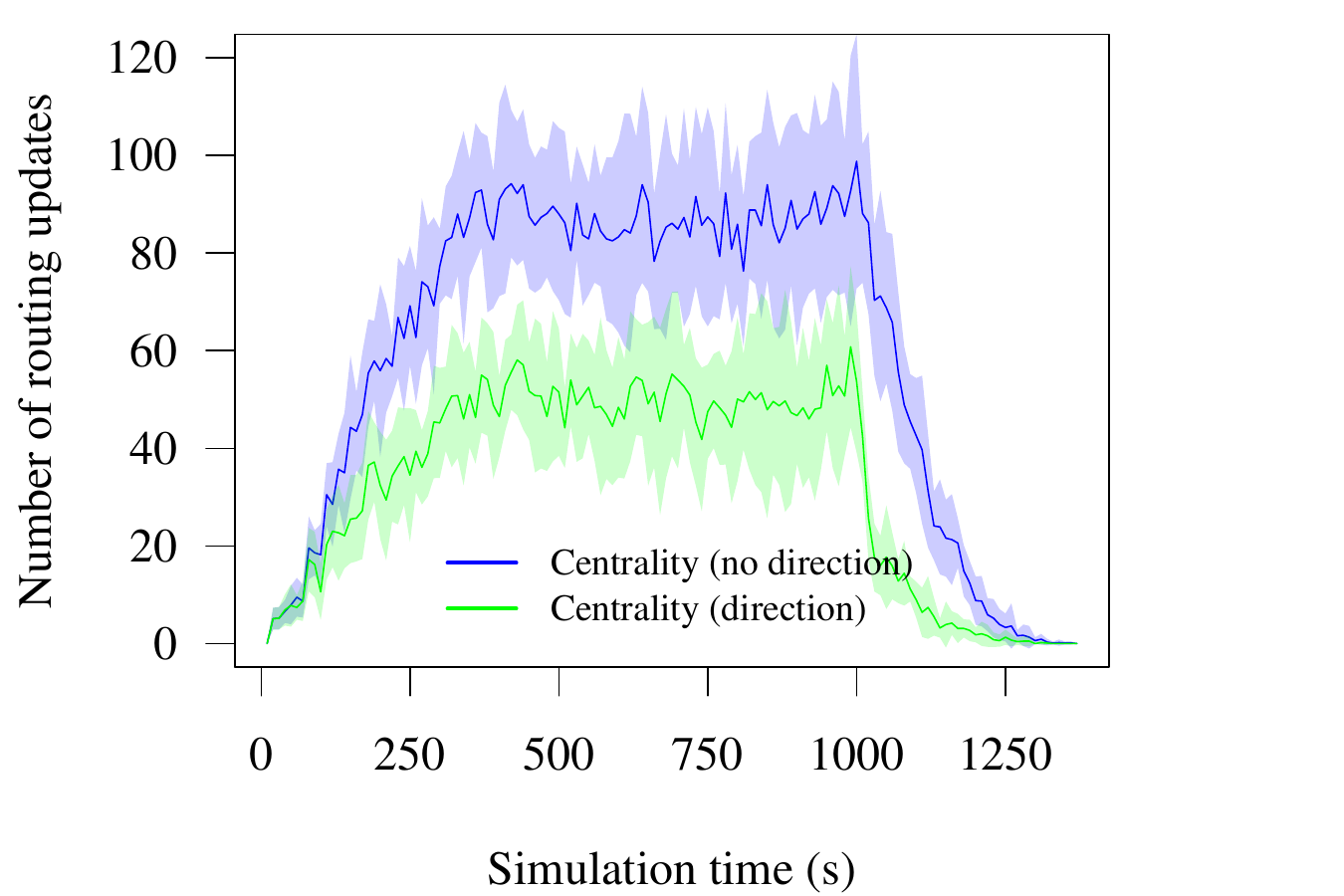}
            \label{fig:cost2}
        }
    \end{center}
    \caption{cost of selecting and maintaining the same aggregation point.}
\end{figure}

Centrality-based algorithm offers a convenient solution for reducing the cost of exchanging control messages.
Figure \ref{fig:cost1} depicts the average and standard deviation of the number of notification messages sent when selecting aggregation points. It generally sends a notification message every time the establishment of an aggregation point. However, we can minimize the number of transmissions by sending a notification message only when a vehicle becomes or ceases to be an aggregation point.
When we consider the direction of vehicles, the average number of notification messages was 18.65, with a range between 10 and 49 during the plateau.
While the approach without direction had an average of 27.68 notification messages, ranging from 19 to 54.
Centrality with direction brought an average reduction of 32.62\% compared to the no-direction approach. This result suggests that network stability contributes to a reduction in the cost of selecting aggregation points.
In addition, Figure \ref{fig:cost2} shows the average and standard deviation of the number of updates in the routing table. Assume that all vehicles maintain a routing table with up-to-date aggregation point information. This information is propagated between neighboring vehicles whenever a new aggregation point is selected. That is, no update occurs when the aggregation point is re-elected. Again, there was a cost reduction when considering the approach with direction. It was possible to reduce the average number of updates in the routing table by 45.39\%. When we use the direction of vehicles, the average number of routing updates is 32.70, with a range between 17 and 95 during the plateau. While the approach without direction had an average of 59.88 routing updates, ranging from 45 to 144. These findings suggest that the proposed algorithm effectively improves the cluster management cost in the two-way roadway scenario.


Compared to the Centrality-based algorithm without direction information, the approach with direction information has an additional computational cost $\mathcal{O}(m)$ to compute the direction of each vehicle and it also reduces the aggregation rate due to the increase in the number of aggregation points. Moreover, without a replacement of aggregation points, there could be a vehicle overload with communication and processing. However, note that prolonging the time of a vehicle acting as an aggregation point improve network stability, reduces the cost of selecting aggregation points, with fewer notification messages transmitted, and decrease the cost of maintaining clusters, requiring fewer updates to update the aggregation point.
Adopting a replacement of aggregation points can contribute to better management of computational resources. This behavior occurs because load distribution among different aggregation points allows the use of processing capacity more uniformly, preventing overload in some nodes and idleness in others. However, evaluating the criteria for choosing the aggregation point is important, as frequent changes increase the cost of forming and maintaining the clusters.

\subsection{Decentralized scenario}

In the decentralized scenario, we chose a sub-map extracted from the central region of Cologne\footnote {https://sumo.dlr.de/docs/Data/Scenarios/TAPASCologne.html}. 
This choice is due to the high cost of the simulations. 
We used traffic traces from \unit[7:00]{am} to \unit[7:10]{am}, with around 323 vehicles distributed over an area of \unit[6.38]{km$^2$}.
We deployed four LTE Base Stations (BS) at different locations to cover all vehicles in the scenario. 

The evaluation is performed through simulations using the SimuLTE v.1.0.1~\citep{virdis} simulator, which implements a detailed LTE stack model and simulates the LTE-Radio Access Network data plane and Evolved Packet Core. 
SimuLTE is an open-source project building on top of OMNeT++\footnote{https://omnetpp.org/}, and it can also work with the Veins framework~\citep{Nardini2017}, which allows simulating the mapped trips from the TAPASCologne project (performed by SUMO) and D2D communications in the vehicular network environment. 
Table \ref{tab:parameters} summarizes the simulation parameters. 

\begin{table}[htb]
    \begin{small}
    \caption{Simulation parameters.}
    \begin{center}
    \begin{tabular}{ll}
        \toprule
        \textbf{Parameters} & \textbf{Values} \\
        \midrule
        Carrier Frequency & \unit[2]{GHz} \\
        Bandwidth         & \unit[5]{MHz} (25 RBs)\\
        Collect Interval  & \unit[5]{s}\\ 
        Delivery Interval & \unit[10]{s}\\ 
        Handover & Enable\\
        Power Transmission & eNodeB: \unit[46]{mW} / UE: \unit[26]{mW} \\
        Antenna Gain & eNodeB: \unit[18]{dBi} / UE: \unit[0]{dbi} \\ 
        Path Loss & ITU-R, Urban Macro Cell model\\
        Thermal Noise & \unit[$-$104.5]{dBm} \\
        \bottomrule
    \end{tabular}
    \label{tab:parameters}
    \end{center}
\end{small}
\end{table}

Besides the aggregation rate and the number of reelections, we used the following ones to evaluate this scenario:
(i) \textit{Cellular throughput}: throughput of transmitted sensory data in the cellular uplink channel. We used this metric to analyze the size of the data reduction in the cellular uplink after data offloading;
(ii) \textit{D2D throughput}: it is the data rate (bits/sec) of the data packets delivered over D2D sidelinks toward all aggregation points considered over the simulation time, i.e., the total number of bits successfully delivered to all the aggregation points divided by the simulation time. We use this metric to analyze the average amount of traffic flowing from one vehicle to its aggregation point in a given period;
(iii) \textit{End-to-end delay}: average delay between a vehicle and the aggregation point. This metric is vital because some sensing applications need a near real-time delivery, but multi-hop communication can generate a considerable delay;
and (iv) \textit{D2D Overhead}: average amount of bytes transmitted over the D2D network when sensing beacons and reservation messages. We use this evaluation to analyze the extra bytes required for data offloading.

\begin{figure}[b]
    \begin{center}
        \includegraphics[width=0.99\linewidth]{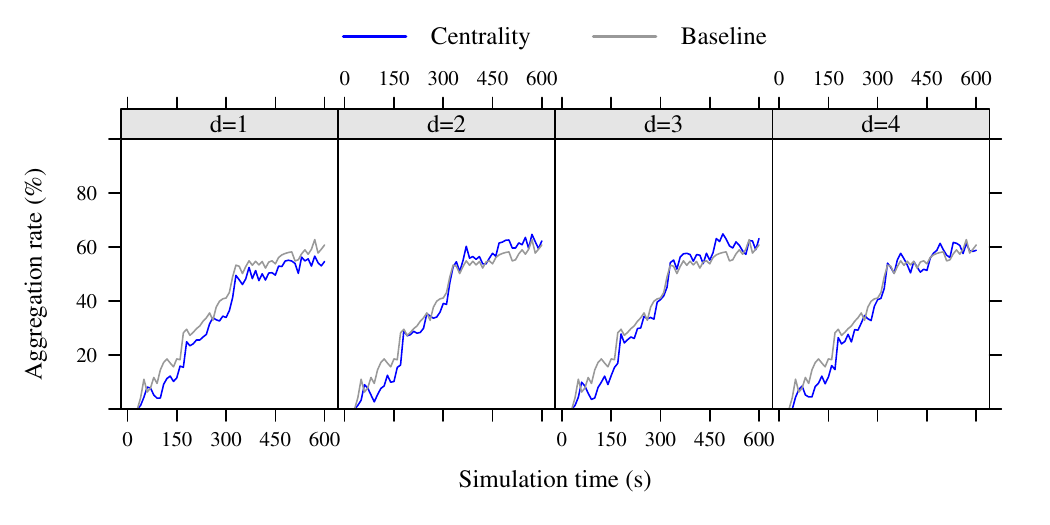}
        \caption{Aggregation rate for single and multi-hop communication.}
        \label{fig:rate1}
    \end{center}
\end{figure}

Figure \ref{fig:rate1} shows the aggregate rate for each value of $d$ and $k = 4$. The single-hop communication scenario shows that the baseline approach is better than the centrality-based algorithm. In contrast with the RB, the centrality-based algorithm models the aggregation capability at each base station so a vehicle can only offload data to aggregation points in the same base station. This strategy reduces the length of the shortest paths in the network. Such behavior can fragment and compromise vehicle connectivity in a sparse network like ours. As we will see below, our algorithm has better cluster stability results than the baseline approach.

From the multi-hop communication view, we improve the aggregation rate by increasing the number of hops. For instance, when $d = 2$, the centrality-based algorithm's aggregation rate was 64.78\%, at best. This scenario represents an 8.78\% increase if we compare it with the value $d = 1$. The results show that the centrality-based is better than the RB after \unit[300]{s} of simulation time.
Although there is an increase in the number of vehicles in the scenario, the average length of the shortest paths is small, around $2.69 \pm{0.81}$ hops. Therefore, the curves show similar performance in multi-hop communication. For $d = 3$ and $d = 4$, the centrality-based increased the aggregation rate of 6.87\% and 4.21\% over the RB algorithm, respectively.
In all cases, we consider $k = 4$. According to previous results, the aggregation rate is better when $k > d$. Consequently, $d = 4$ presents a lower performance than $d = 3$.

\begin{figure}[htbp]
    \centering
    \subfigure[Throughput in the cellular network.]{
    \includegraphics[width=.42\textwidth]{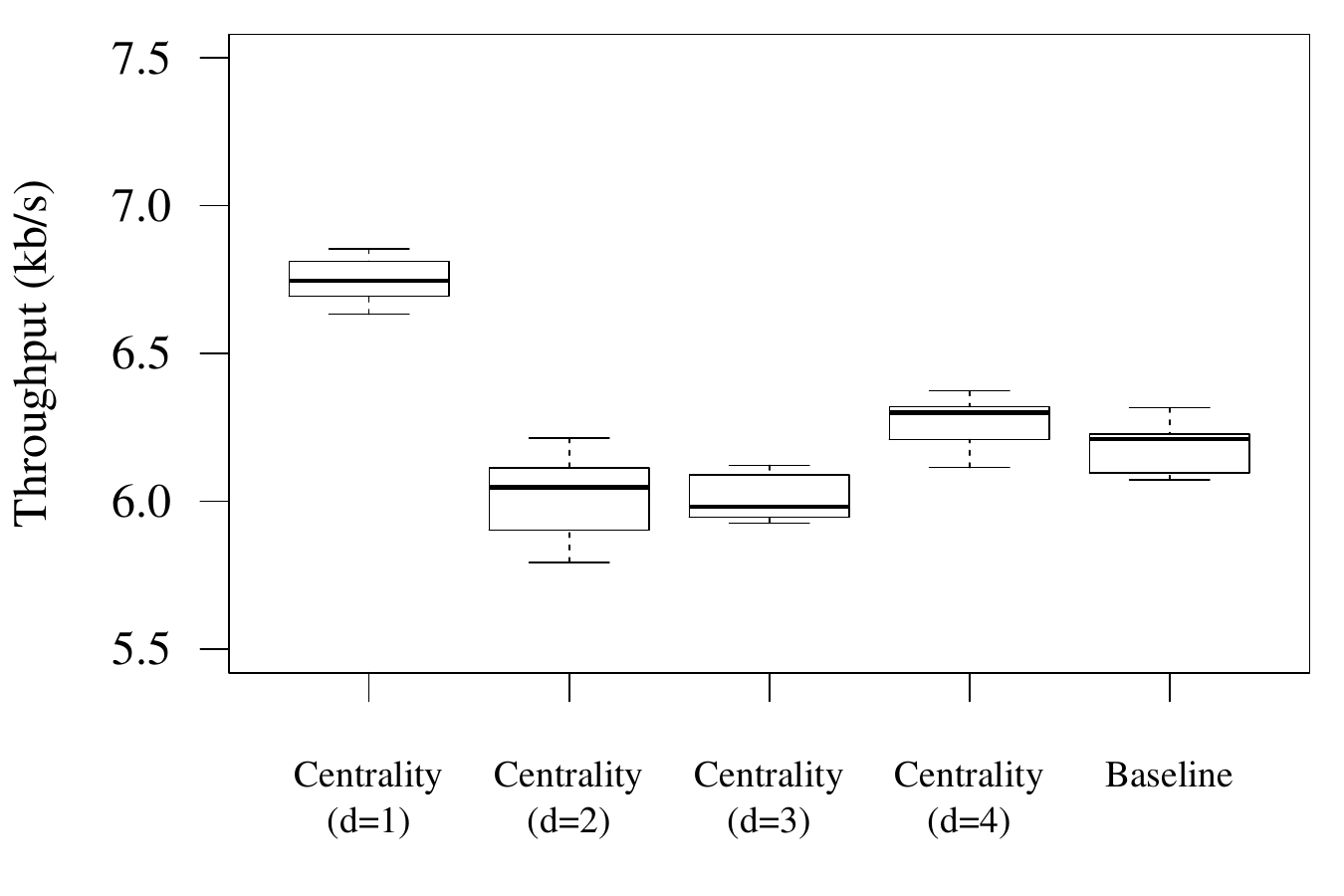}
    \label{fig:throughput1}
    }
    \subfigure[Throughput in the D2D communication.]{
        \includegraphics[width=.42\textwidth]{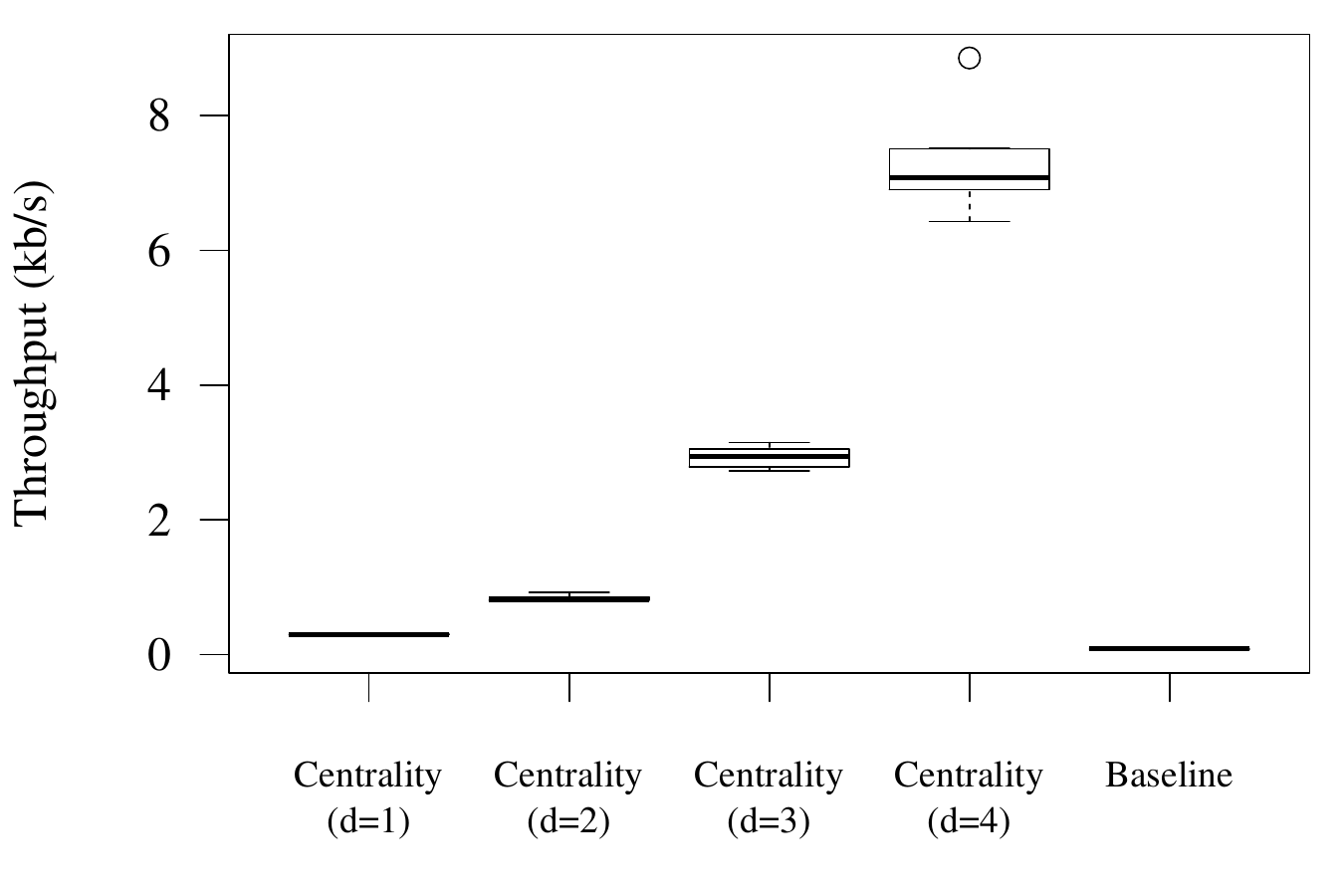}
        \label{fig:throughput2}
    }
    \subfigure[End-to-end delay.]{
    \includegraphics[width=.42\textwidth]{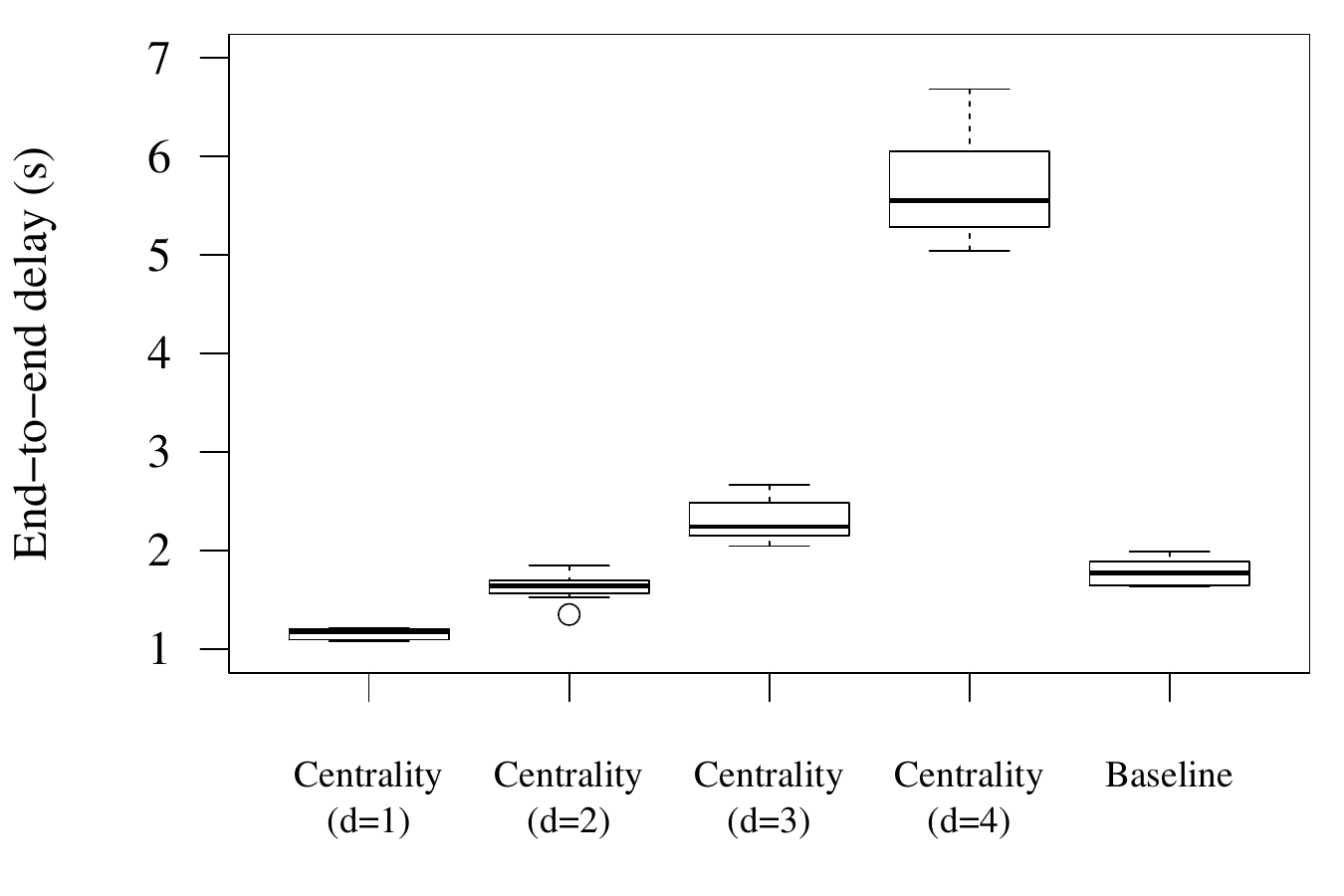}
    \label{fig:delay}
    }
    \subfigure[Number of re-elections.]{
    \includegraphics[width=.42\textwidth]{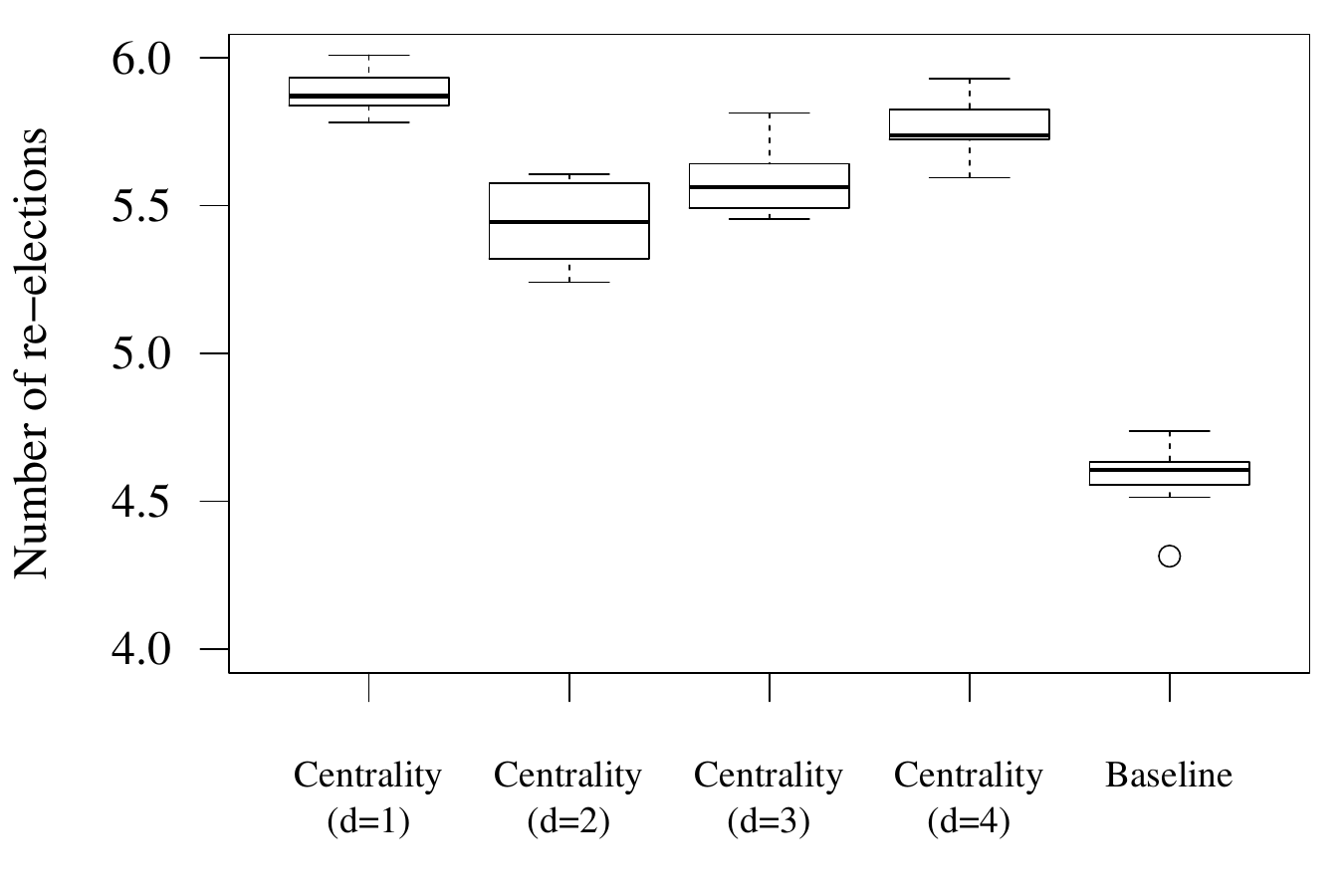}
    \label{fig:reelection}
    }
    \caption{evaluation of data transfer in the decentralized scenario.}
\end{figure}

Figure \ref{fig:throughput1} shows the throughput in the cellular network. We compute it with the data transmitted in the uplink of the four base stations. The upload without an offloading system reached a throughput of up to \unit[13.23]{kb/s}. These results show that offloading reduces up to 52.32\% of the throughput compared to traditional upload.
Although the experiment does not reproduce a scenario of intense overload on the cellular network, the proposed solution can reduce the uplink overload, as observed in the results. The application can also process the data at the edge of the cellular network on a MEC platform. It could also reduce the overload in the core network to facilitate further access to data available for real-time or location-based applications.


Figure \ref{fig:throughput2} shows the D2D communication's throughput. In the case of D2D communications, neighbor discovery is an essential service. We use the direct discovery method, where each vehicle announces its presence through periodic signaling messages (beacons) broadcast. Results show that the D2D channel utilization is more significant in the centrality-based algorithm due to signaling overhead and the multi-hop transmissions. Multi-hop communication can find even more minor aggregation points, but several practical challenges involve multi-hop communication. For example, we can overload the complementary network without an efficient data collection strategy.
Although the applications can offload the data over unlicensed frequency spectrums, such as Wi-Fi networks, there is still a cost since a large amount of data could compromise the capacity of the D2D communication channel. 
In contrast, the RB algorithm may suffer synchronization overhead, but this work did not examine it. A complete simulation could analyze the synchronization in detail.

Figure \ref{fig:delay} analyzes the impact of the number of hops on the end-to-end delay. For example, if the aggregation point is far, it receives the data with a high end-to-end delay.
The centrality-based algorithm uses closeness centrality to select aggregation points that reduce the transmission time according to the number of hops. Our study reveals that the proximity relationship between vehicles can reduce the end-to-end delay for local data aggregation. As a result, sensor data takes, on average, 1.18 seconds (better than the baseline) to reach its aggregation point via single-hop communication.
There is a trade-off between the aggregation rate and end-to-end delay.
Multi-hop communication can improve the aggregation rate, increasing the end-to-end delay. This issue can degrade the QoS performance of real-time applications, such as online sensing applications.

Our analysis also reveals that the centrality-based algorithm produces better stable clusters than the baseline. Figure \ref{fig:reelection} shows stability results for both solutions. Stable clusters tend to reelect the exact vehicle as an aggregation point. Periodic changes in aggregation points can increase network overhead due to the routing information. In contrast, the reelection of aggregation points can minimize network traffic. Clusters formed by the centrality-based algorithm are more stable than those created by the RB algorithm in single and multi-hop communication. 
We observe the reelection of aggregation points by an average between 6 and 4 times, but there are aggregation points reelected up to 43 times by the centrality-based algorithm. The maximum number of reelections was 33 times for the RB algorithm.


The results show that the $k$-closeness centrality can reduce the average path length between the aggregation points and other vehicles. Therefore, nodes with higher centrality are ideal for data collection with fewer rebroadcasts and latency.
Table \ref{tab:b2} shows the results of a two-tailed Wilcoxon signed-rank test, comparing mean values for each variable in the single-hop communication of the centrality-based and the RB algorithm. The results suggest that all pairs of samples have statistical differences. The RB algorithm is better than the centrality-based algorithm in aggregation rate but presents longer delays and less stability. Nevertheless, we can solve the aggregation rate problem by increasing the number of hops. 

\begin{table}[htbp]
    \renewcommand{\arraystretch}{1.3}
    \scriptsize
    \caption{Wilcoxon signed-rank test for paired samples (two-tailed).} 
    \begin{center}
\begin{adjustbox}{width=\linewidth,keepaspectratio,center=\linewidth}
    \begin{tabular}{lccccc}
        \toprule[1.0pt]
        \multirow{2}{*}{Centrality - RB} & \multirow{2}{*}{(Pseudo) Median} & \multirow{2}{*}{Std. Deviation} &  \multicolumn{2}{c}{Confidence interval (95\%)} & \multirow{2}{*}{p-value}\\
        \cline{4-5}
           &  & & Lower & Upper & \\
        \midrule[1.0pt]
          Aggregation rate & -4.685 & 2.595 & -5.257 &  -4.139 & 1.073e-10 \\
          Cellular throughput & 0.069 & 0.008 & 0.062 & 0.074 & 0.006\\ 
          D2D throughput & 0.027 & 0.001 & 0.026 & 0.027 & 0.002\\
          Transmission delay & -0.622 & 0.139 & -0.739 & -0.506 & 0.002\\
          Number of re-elections & 1.281 & 0.106 & 1.238 & 1.383 & 0.002\\
        \bottomrule[1.0pt]
    \end{tabular}
    \label{tab:b2}
\end{adjustbox}
    \end{center}
\end{table}

If we only consider the time range of the highest vehicle density (after 300 seconds of the simulation), the paired test (one-tailed) reports a $p$-value of 7.255e-05 when the number of hops equals 2. This result suggests that the centrality-based algorithm is statistically better than the baseline in this specific scenario. However, there are limitations to our approach. The algorithm performs poorly when the network is highly fragmented, and the shortest paths must be better defined. Besides, multi-hop communication generates considerable overhead in the D2D network, so we must adjust the parameters appropriately to find a trade-off between the transmission cost on the sidelink and the aggregation rate.

\begin{figure}[t]
    \centering
    \includegraphics[width=.8\linewidth]{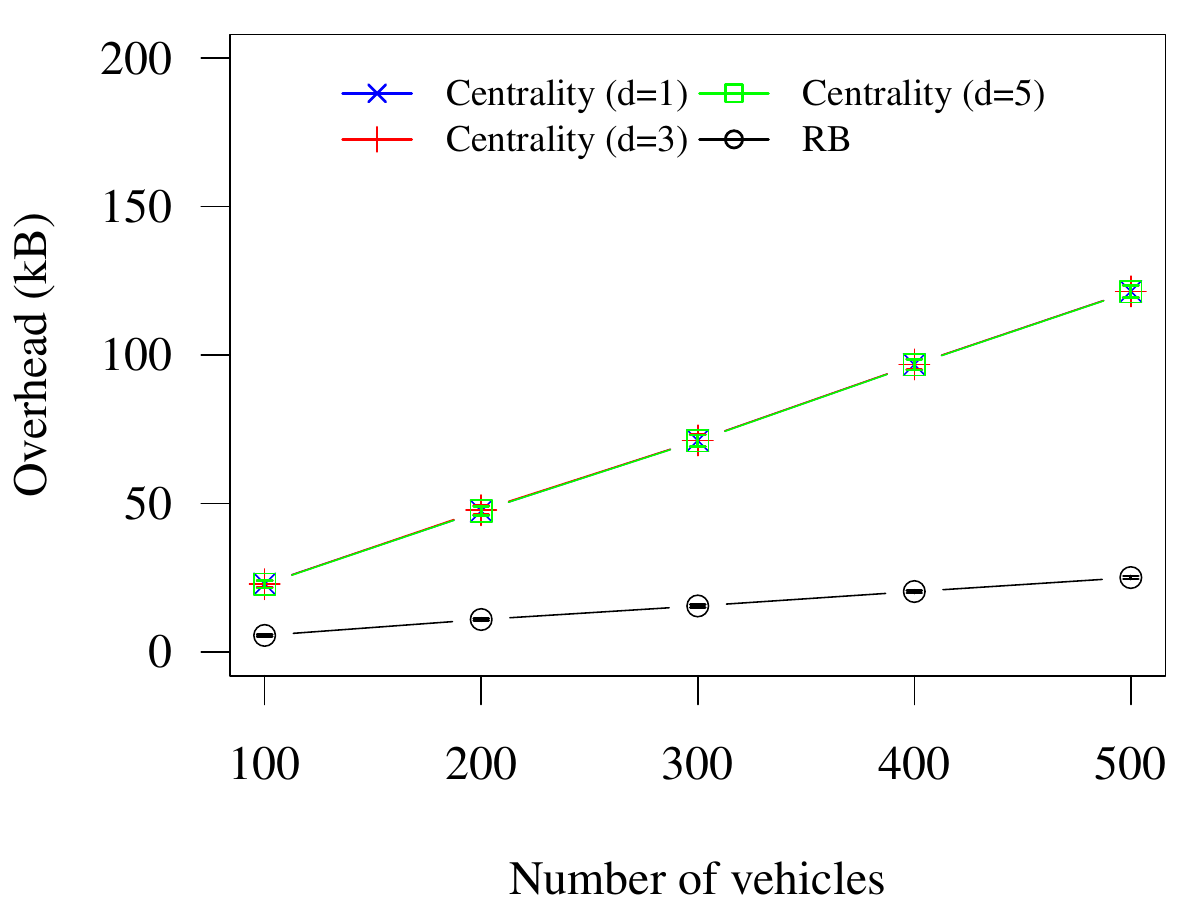}
    \caption{D2D Overhead versus the number of vehicles.}
    \label{fig:overhead}
\end{figure}

Finally, we analyzed the impact of vehicular density on overhead and delay in the D2D network.
We use the \textbf{SUMO random trips tool} to generate a set of random trips and show the impact of the density of vehicles in the Cologne road network (central area). We perform the network simulation using SimuLTE v.1.0.1 simulator -- we use the same simulation parameters (see Table~\ref{tab:parameters}). Five different setups were analyzed: 100, 200, 300, 400, and 500. Figure~\ref{fig:overhead} presents the average overhead in the D2D network communication in terms of the transmitted beacon and reservation messages, together with the standard deviation. The centrality-based algorithm suffers from a higher overhead than the RB algorithm, where each vehicle transmits periodic messages to keep the neighborhood structure updated. However, there is no significant difference between single-hop and multi-hop communication since it sends only one-hop beacon messages.
In the RB algorithm, dominated vehicles do not transmit reservation messages. Therefore, the overhead is proportional to the number of transmitted reservation messages, which tends to decrease with the number of aggregation points. As previously reported, there is overhead due to synchronization in the RB algorithm. However, we do not evaluate this aspect in this work.

\begin{figure}[htb]
    \centering
    \includegraphics[width=.8\linewidth]{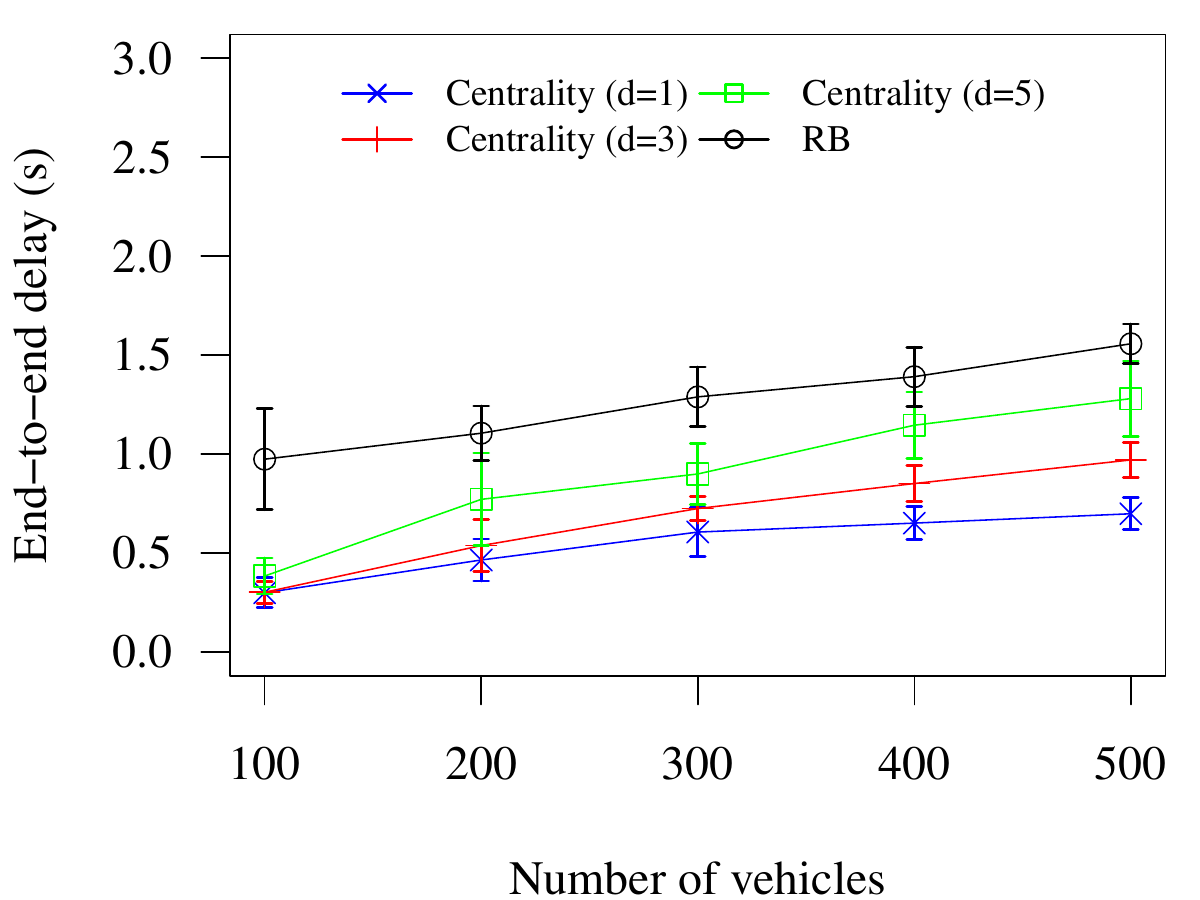}
    \caption{End-to-end delay versus the number of vehicles.}
    \label{fig:delay2}
\end{figure}

In addition, Figure~\ref{fig:delay2} presents the average end-to-end delay and standard deviation. The centrality-based algorithm performed better in terms of end-to-end delay. Again, there is a longer delay when multi-hop communication is enabled. This delay increases with the number of vehicles in the scenario because more data flows in the D2D network, resulting in a long time to process the packets. These results highlight how the centrality measure can reduce the delay between sensor vehicles and aggregation points. While the RB chooses the aggregation points randomly, the centrality-based algorithm chooses them based on their proximity to the other vehicles.

The decentralized scenario showed a more realistic evaluation with four base stations distributed throughout the communication area. We perform the aggregation point selection at each base station in a decentralized manner according to the local view of the system. Due to the high simulation cost, we reduced the scenario for simulation, resulting in a highly fragmented network. The results suggest that the centrality-based algorithm could perform better in networks with this topology. Therefore, we have included the centralized scenario that comprises a large-scale urban area with all vehicles associated with the same base station. This scenario simulated a dense and highly connected network, but because of the high simulation cost, we do not use SimuLTE in this case. Finally, we performed a high-level evaluation, analyzing the multi-hop communication and the impact of the vehicle directions. The results showed a significant gain in multi-hop communication when the network is densely connected, while vehicles' direction contributes to the network's fragmentation.

\section{Conclusion}\label{sec:conclusion}

In this work, we proposed an algorithm for selecting aggregation points to perform data offloading in a vehicular sensor network. The algorithm uses a centrality metric to find the most compatible vehicles in a data offload scenario. The proposed solution is a decentralized heuristic and does not need an infrastructure complementary to the cellular network.
In addition, our heuristic does not require synchronized transmissions between vehicles and is scalable to support different traffic conditions. The centrality metric only considers local calculations, resulting in more stable clusters and minor end-to-end delays. Results showed that we could use neighborhoods of different sizes to estimate the centrality measure adaptively, so it was possible to obtain a high aggregation rate even in scenarios with high vehicular density, where we reduced the diameter of the calculation.
In future work, we will consider a more robust evaluation with real urban monitoring applications, in which we can explore different conditions of data aggregation and QoS requirements.

\bibliographystyle{plain}

\bibliography{ref}

\end{document}